\documentclass[journal]{IEEEtran}
\IEEEoverridecommandlockouts
\usepackage{amsmath,amssymb,amsfonts}
\usepackage{graphicx, hyperref, url}
\usepackage{textcomp}
\usepackage{csquotes}
\usepackage{xcolor,soul}
\usepackage{adjustbox}
\usepackage{hyperref}
\usepackage{multirow}
\usepackage{titlesec}
\usepackage{subcaption}
\usepackage{makecell}
\usepackage{amsfonts}
\usepackage{algorithm}
\usepackage{algpseudocode}

\usepackage[numbers,sort&compress]{natbib}

\usepackage{etoolbox}

\definecolor{columbiablue}{rgb}{0.61, 0.87, 1.0}
\sethlcolor{white}

\usepackage[most]{tcolorbox}

\definecolor{changedbg}{RGB}{255,248,210} % light yellow

\newtcolorbox{changedsection}{
    enhanced,
    breakable,
    colback=changedbg,
    colframe=changedbg,
    boxrule=0pt,
    arc=0pt,
    left=3pt,
    right=3pt,
    top=3pt,
    bottom=3pt,
    before skip=5pt,
    after skip=5pt
}

\titlespacing{\subsubsection}{0pt}{0pt}{1pt}

\usepackage{amsmath}

\usepackage{mathtools}
\usepackage[colorinlistoftodos,prependcaption,textsize=tiny]{todonotes}
\usepackage{gensymb}
\usepackage{multirow}
\usepackage{orcidlink}
\usepackage{parskip}

\usepackage{titlesec}

\usepackage{amsthm}
\theoremstyle{definition}
\newtheorem{theorem}{Theorem}
\newtheorem{lemma}{Lemma}

\renewcommand{\thesubsection}{\Alph{subsection}}
\renewcommand{\thesubsubsection}{\thesubsection.\Roman{subsubsection}}
\usepackage{titlesec}
\titleformat{\subsubsection}
  {\normalfont\normalsize\bfseries}
  {\thesubsubsection}
  {1em}
  {}

\begin{document}

\title{Multi-Scaled Unscented Kalman Filter}
\author{Amit Levy \orcidlink{0009-0009-8703-5807},
        and~Itzik~Klein \orcidlink{0000-0001-7846-0654}
% <-this % stops a space
\thanks{A. Levy and I. Klein are with the Hatter Department of Marine Technologies, Charney School of Marine Sciences, University of Haifa, Israel.\\ Corresponding author: alevy02@campus.haifa.ac.il}% <-this % stops a space
}

\maketitle
% ---------------------Article -------------------------------------
% ------------------------ Abstract
\begin{abstract}
The unscented Kalman filter (UKF) is a commonly used algorithm capable of estimating the states of nonlinear dynamic systems. It chooses a set of sigma points that capture the states posterior mean and covariance of the nonlinear system. The filter is based on the scaled unscented transform, where the scaling parameters impact the spreading of the sigma points, determining the estimated model capturing. In its current form, the UKF employs a single set of scaling parameters shared by all sigma points. Because states in multi-dimensional models often exhibit substantially different behaviors, this imposes a critical limitation: the standard UKF parameters cannot be tuned to extend the spread for one dimension while reducing it for another. To bridge this gap, we propose the multi-scaled UKF to enable spreading differently per state, fitting better the underlying model behavior, while maintaining the key properties of the sigma points and UKF. A rigorous mathematical foundation is provided, introducing the theoretical approach to multi-scaling. The benefits of this approach are demonstrated through two distinct nonlinear dynamic systems. Consequently, our multi-scaled UKF captures the nonlinear behavior of multi-dimensional states more effectively, leading to improved performance and estimation accuracy.
\end{abstract}

% ------------------------ Introduction
\section{Introduction}\label{intro_sec}
The linear Kalman filter (KF) \cite{Kalman1960} is an optimal estimator considering minimum mean squared error, designed to estimate the states of linear dynamic systems assuming Gaussian state distributions and Gaussian noise. However, in many practical systems, the dynamic model, the measurement model, or both are nonlinear. Therefore, the extended Kalman filter (EKF) was introduced to address such nonlinearities by linearizing the system and measurement models, using first-order partial derivatives (Jacobian matrices), producing a local linear approximation to which the standard KF can be applied \cite{barShalomappandtrack2004}. The linearization procedure introduces errors in the posterior mean and covariance, which may result in suboptimal performance and in some cases a filter divergence \cite{WanUkfnonlin2000}. To address this limitation, Julier and Uhlmann proposed the unscented Kalman filter (UKF) \cite{JUlUKF1997}, based on the unscented transform (UT). The underlying principle is that it is generally easier to approximate a probability distribution than a nonlinear function. The UKF employs a selected set of sigma points to capture the system state mean and covariance. When these sigma points are propagated through the nonlinear system, the posterior mean and covariance are accurately approximated up to the second order and up to the third order in the case of a Gaussian distribution.\\
Numerous studies have compared the performance of the EKF and the UKF, showing the superiority of the latter. For example, the advantage of the UKF for the problem of polar-to-Cartesian coordinate transformation was demonstrated in \cite{JUlUKF2004}. Analysis and comparison between the EKF and UKF was introduced in \cite{kandepu2008applying}, concluding that UKF is better in terms of performance, robustness and convergence speed. Further, the usage of UKF for dynamic alignment of low cost inertial measurement unit, demonstrated that the UKF allows large initial attitude errors and the transition from small to large attitude errors can be made smoothly, without the need to change the model \cite{shin2004unscented}.\\
Beyond theoretical developments, the UKF has been extensively implemented across a wide range of practical applications. In \cite{hao2007comparison}, the Rao-Blackwellised additive UKF is proposed to mitigate a dynamic system with additive noise. The UKF in \cite{xiong2009modified} was applied to determine autonomously Earth's satellites orbit with a unique modification to the noise covariance matrix. The UKF proposed in \cite{lv2015sensorless} estimates the speed and the rotor position of a brushless DC motor using voltages and currents measurements only. Round-Robin Protocol-based UKF and Weighted Try-Once-Discard Protocol-based UKF are proposed in \cite{ukf_for_stochastic_2020} for a class of general nonlinear systems with stochastic uncertainties. A fault detection method using the UKF is proposed in \cite{faultDetectUkf2025} for state estimation in nonlinear stochastic systems subject to strong noise and communication protocols. A multivariate Gaussian process and the UKF are integrated into a learning-based model predictive control framework in \cite{MGPUkf2026}, for trajectory tracking of autonomous surface vehicles. The UKF's utility in autonomous underwater vehicles navigation is demonstrated in \cite{anukf2026}, where a neural network has been utilized to estimate dynamically the process noise covariance matrix. This estimation is critical as it dictates the spread of sigma points via its impact on the state covariance matrix. Subsequently, a UKF featuring a neural network backbone was proposed to estimate the process and the measurement noise covariances from raw data of inertial sensors and GNSS signals, improving the position accuracy of an unmanned ground vehicle navigation solution \cite{versano2026hybrid}.\\
Several studies address the modification, spreading, and propagation of UKF sigma points, highlighting their critical impact on performance and accuracy. In \cite{levy2025unscented}, a propagation method was introduced by coupling the error-state model with the nominal system model, thereby enhancing AUV navigation across various operational scenarios. A reduction in the number of the required sigma points has been analyzed and validated using a highly nonlinear univariate nonstationary growth model (UNGM) \cite{optselectsigmapoints2011}. In \cite{systematicUKF2015}, a methodology for the UKF theory was introduced by reconsidering the representation of the sigma point set and its size, providing a consistent adjustment utilizing a minimum number of sigma points. Conversely, \cite{highOrderUkf2003} and \cite{ModUkf2018} introduced modifications that enlarge the sigma point set to minimize higher-order errors. A generalized unscented transform is proposed in \cite{ebeigbe2025genUT}, where the selected sigma points and their associated weights reproduce the input mean and covariance together with selected third and fourth order moments, making it suitable when these moments have a dominant influence on the propagated statistics. Shortly after introducing the UT, Julier \cite{scaledUT2002} introduced the scaled unscented transform, which employs an adjustable parameter to control sigma point spread and better capture the estimated model features. This technique reduces the dependence between state dimension and spreading radius, ultimately improving the accuracy of mean and covariance estimations. The scaling factor's influence on UT behavior was further analyzed in \cite{scaledUTAnalysis2012} using two nonlinear models. Similarly, \cite{adaptScaleUKF2014} investigated the selection of the scaling parameter, demonstrating its importance in a bearings-only tracking problem. In \cite{scalingParmUkf2018}, an effective UKF featuring self-adaptive scaling parameters has been introduced, showing improved performance on both the UNGM and a strongly nonlinear three-dimensional model.\\
In numerous dynamic systems, the nonlinear behavior of the states is not uniform, thus, the sigma point spread should accommodate tuning differently per state behavior. While the UKF allows for the expansion or contraction of sigma point spread to achieve a better fit of the estimated model, these adjustments apply uniformly across all states. Consequently, the parameters cannot be tuned to extend certain sigma points while reducing others. Since individual states may exhibit substantially different behaviors, this disparity poses a limitation on the UKF. This limitation is particularly relevant in industrial control and monitoring applications, where multi-dimensional states routinely combine slow, well-behaved dynamics with fast or highly nonlinear ones. Examples include multi-axis servo and positioning systems such as pan–tilt platforms and azimuth–elevation drives, industrial process control loops with coupled temperature and condition-monitoring systems that fuse vibration and rotational-speed measurements. In such systems, forcing a single scaling set onto all states can degrade the estimation accuracy of the more sensitive state to accommodate the other, directly affecting control performance and fault-detection reliability. A filter that can tune its sigma-point spread independently per state offers practical value beyond the theoretical gain, without requiring any change to the underlying filter architecture or added computational burden. To address this gap, we propose the multi-scaled UKF (MS-UKF). In this manner, we control the spreading differently per state and better capture the underlying multi-dimension nonlinear model, while maintaining the key properties of the sigma points. That is, the MS-UKF improves performance and accuracy compared to the standard UKF in terms of mean and covariance estimations. Our MS-UKF is grounded in a rigorous mathematical foundation, introducing a theoretical approach to multi-scaling.\\
The main contributions of the paper are:
\begin{enumerate}
    \item We develop a modification to the UKF that extends the control of the sigma point spread for multi-dimensional nonlinear models, introducing a multi-scaled set that better captures the estimated model in terms of mean and covariance, improving performance and filter behavior.
    \item We establish a robust mathematical foundation for the multi-scaling approach, complemented by a detailed theoretical analysis.
\end{enumerate}
We demonstrated our approach on two different nonlinear dynamic systems, showing the importance of the enhanced modification on the true
behavior of the filter. Furthermore, as the MS-UKF extends the UKF scaling method, modifications
applied to the UKF can be applied almost transparently to the MS-UKF, utilizing the extended set of scaling parameters to improve the filter performance when trying to estimate a multidimensional nonlinear model.\\
The rest of the paper is organized as follows: Section \ref{prob_form_sec} formulates the problem. Section \ref{prop_appr} introduces our proposed approach, introducing the MS-UKF modified scaling system. Section \ref{res_sec} presents the results, comparing MS-UKF and the standard UKF on two nonlinear dynamic systems. Finally, Section \ref{conc_sec} provides the conclusions of this work.

% ------------------------ Problem Formulation
\section{Problem Formulation}\label{prob_form_sec}
\subsection{The UKF algorithm}\label{ukf_std_alg}
Consider the following dynamic system:
\begin{equation}\label{general_dynamic_model_eq}
    \centering
    \boldsymbol{x}_{k+1} = {f}(\boldsymbol{x}_{k}, \boldsymbol{\omega}_{k})
\end{equation}
where $\boldsymbol{x}_{k}$ is the unobserved state vector at step ${k}$, ${f}$ is the dynamic model mapping, and ${\boldsymbol{\omega}}$ is the system process noise.\\
The measurement model is:
\begin{equation}\label{general_meas_model_eq}
    \centering
    \boldsymbol{z}_{k+1} = {h}(\boldsymbol{x}_{k+1}, \boldsymbol{\nu}_{k+1})
\end{equation}
where, ${h}$ is the observation model mapping, $\boldsymbol{\nu}$ is the measurement noise, and $\boldsymbol{z}_{k+1}$ is the observed signal at step ${k+1}$. \\
Let $\hat{\boldsymbol{x}}_{{k}{|}{k}} \in \mathbb{R}^n$ be the estimated state vector at time step $k$ (left index of $k|k$) given $k$ measurements (right index of $k|k$) and $\boldsymbol{\mathbf{P}}_{{k}{|}{k}} \in \mathbb{R}^{n \times n}$ be the associated estimated covariance matrix at time step $k$. As $\boldsymbol{\mathbf{P}}_{{k}{|}{k}}$ is positive definite, it can be factored by the Cholesky decomposition as follows:
\begin{equation}\label{cholesky_eq}
    \centering
    \boldsymbol{\mathbf{P}}_{{k}{|}{k}}=\sqrt{\boldsymbol{\mathbf{P}}_{{k}{|}{k}}}\sqrt{\boldsymbol{\mathbf{P}}_{{k}{|}{k}}}^{T}
\end{equation}

where $\sqrt{\boldsymbol{\mathbf{P}}_{{k}{|}{k}}}_i$ denotes the ${i}$-th row of the matrix $\sqrt{\boldsymbol{\mathbf{P}}_{{k}{|}{k}}}$. \\
The scaling parameter, $\lambda$ is defined by:
\begin{equation}\label{lambda_std_eq}
    \centering
    {\lambda=\alpha^2(n+\kappa)-n} 
\end{equation}
where $\alpha$ determines the spread of the sigma points around $\hat{\boldsymbol{x}}_{{k}{|}{k}}$. $\kappa$ is a secondary scaling parameter, and $\beta$ is used to incorporate prior knowledge of the distribution of $\boldsymbol{x}$ (for a Gaussian distribution it is set to 2).\\
The compensation scaling factor $\gamma$~\cite{WanUkfnonlin2000}:
\begin{equation}\label{gamma_std_eq}
    \centering
    \gamma = 1 - \alpha^2 + \beta,
\end{equation}
is used in the calculation of the covariance matrix.\\
Following the initialization step (step 1), the algorithm enters an iterative cycle consisting of three main stages. While steps 2 and 3 are performed regularly, step 4 is executed only upon the availability of measurements. These steps are detailed below:
\begin{enumerate}
    \item \textbf{Initialization}:
        Given the initializing random ${n}$ state vector ${x}_{0}$ with known mean and covariance, set
        \begin{equation}\label{init_x_eq}
            \centering
            \hat{\boldsymbol{x}}_{0} = {E[\boldsymbol{x}_{0}]}
        \end{equation}
        \begin{equation}\label{init_p_eq}
            \centering
            \boldsymbol{\mathbf{P}}_{0|0} = {E[ (\boldsymbol{x}_{0}-\hat{\boldsymbol{x}}_{0})(\boldsymbol{x}_{0}-\hat{\boldsymbol{x}}_{0})^{T} ]}
        \end{equation}
        The weights for the mean calculation (marked by uppercase letter m) and the covariance calculation (marked by uppercase letter c) are:
         \begin{equation}\label{w_m_0_eq}
            \centering
            w_{0}^{m} = \frac{\lambda}{(n+\lambda)}
        \end{equation}
         \begin{equation}\label{w_c_0_eq}
            \centering
            w_{0}^{c} = \frac{\lambda}{(n+\lambda)} + \gamma
        \end{equation}
         \begin{equation}\label{w_i_eq}
            \centering
            w_{i}^{m} = w_{i}^{c} = \frac{1}{2(n+\lambda)}, {i=1,...,2n}
        \end{equation}
            
        \item \textbf{Sigma points update}: The ${2n+1}$ sigma points are defined as follows:
         \begin{equation}\label{sp_0_eq}
            \centering
            \boldsymbol{x}_{0,{{k}{|}{k}}} = \hat{\boldsymbol{x}}_{{k}{|}{k}}
        \end{equation}
         \begin{equation}\label{sp_i_eq}
            \centering
            \begin{aligned}
            \boldsymbol{x}_{i,{{k}{|}{k}}} =& \hat{\boldsymbol{x}}_{{k}{|}{k}} + \left(\sqrt{({n}+\lambda)\boldsymbol{\mathbf{P}}_{{k}{|}{k}}}\right)_{i},
            \\& {{i}= {1},...,{n}}
            \end{aligned}
       \end{equation}
        \begin{equation}\label{sp_n_i_eq}
            \begin{aligned}
            \centering
            \boldsymbol{x}_{i,{{k}{|}{k}}} =& \hat{\boldsymbol{x}}_{{k}{|}{k}} - \left(\sqrt{({n}+\lambda)\boldsymbol{\mathbf{P}}_{{k}{|}{k}}}\right)_{i-n}, \\
            & {{i}= {n+1},...,{2n}}
            \end{aligned}
        \end{equation}
        
    \item \textbf{Time update}:
        The sigma points are propagated through the nonlinear dynamics, after which the predicted mean and covariance are computed as follows:
        \begin{equation}\label{tu_x_eq}
            \centering
            \boldsymbol{x}_{i,{{k+1}{|}{k}}} = f(\boldsymbol{x}_{i,{{k+1}{|}{k}}}), i=0,...,2n
        \end{equation}
        \begin{equation}\label{tu_x_mean_eq}
            \centering
            \hat{\boldsymbol{x}}_{{k+1}{|}{k}} = \sum_{i=0}^{2n} w_{i}^{m} \boldsymbol{x}_{i,{{k+1}{|}{k}}}
        \end{equation}
        \begin{equation}\label{tu_p_eq}
            \centering
            \boldsymbol{\mathbf{P}}_{{k+1}{|}{k}}=
            \sum_{i=0}^{2n}w_{i}^{c}{\boldsymbol{\delta x}_{i,{{k+1}{|}{k}}} \cdot {\boldsymbol{\delta x}_{i,{{k+1}{|}{k}}}}^T} + \boldsymbol{\mathbf{Q}}_{k}
        \end{equation}
        where 
        \begin{equation}\label{eq:newx}
        \boldsymbol{\delta x}_{i,{{k+1}{|}{k}}} = \boldsymbol{x}_{i,{{k+1}{|}{k}}}-\hat{\boldsymbol{x}}_{{k+1}{|}{k}}
        \end{equation}
        and $\boldsymbol{\mathbf{Q}}_{k}$ is the covariance process noise matrix at step ${k}$

    \item \textbf{Measurement update}:
        The sigma points are propagated through the observation model to predict the measurement.
        For each sigma point, the estimated measurement is:
        \begin{equation}\label{z_i_eq}
            \centering
            \boldsymbol{z}_{i,{{k+1}{|}{k}}} = h(\boldsymbol{x}_{i,{{k+1}{|}{k}}}), i=0,...,2n ,
        \end{equation}
        and the mean estimated measurement is:
        \begin{equation}\label{z_mean_eq}
            \centering
            \hat{\boldsymbol{z}}_{{k+1}{|}{k}} = \sum_{i=0}^{2n}{w}_{i}^{m}\boldsymbol{z}_{i,{{k+1}{|}{k}}}
        \end{equation}
        The measurement covariance matrix is computed by:
        \begin{equation}\label{s_eq}
            \centering
            \boldsymbol{\mathbf{S}}_{k+1} = \sum_{i=0}^{2n}{w}_{i}^{c}\boldsymbol{\delta z}_{i,{{k+1}{|}{k}}}\cdot {\boldsymbol{\delta z}_{i,{{k+1}{|}{k}}}}^T+\boldsymbol{\mathbf{R}}_k
        \end{equation}
        where 
        \begin{equation}\label{eq:newz}
        \boldsymbol{\delta z}_{i,{{k+1}{|}{k}}} = \boldsymbol{z}_{i,{{k+1}{|}{k}}} - \hat{\boldsymbol{z}}_{{k+1}{|}{k}}
        \end{equation}
        and $\boldsymbol{\mathbf{R}}_k$ is the measurement noise matrix at step $k$.
        
        The cross-covariance matrix is calculated using \eqref{eq:newx} and \eqref{eq:newz}:
        \begin{equation}\label{p_cross_cov_eq}
            \centering
            \boldsymbol{\mathbf{P}}_{k+1}^{x,z} = \sum_{i=0}^{2n}{w}_{i}^{c}\boldsymbol{\delta x}_{i,{{k+1}{|} {k}}}\cdot \boldsymbol{\delta z}_{i,{{k+1}{|}{k}}}^T
        \end{equation}
        %where $\delta\boldsymbol{x}_{i,{{k+1}{|}{k}}} = \boldsymbol{x}_{i,{{k+1}{|}{k}}}-\hat{\boldsymbol{x}}_{{k+1}{|}{k}}$ and $\delta\boldsymbol{z}_{i,{{k+1}{|}{k}}} = \boldsymbol{z}_{i,{{k+1}{|}{k}}} - \hat{\boldsymbol{z}}_{{k+1}{|}{k}}$.\\
        Using \eqref{s_eq} and \eqref{p_cross_cov_eq} the Kalman gain is calculated by:
        \begin{equation}\label{k_eq}
            \centering
            \boldsymbol{\mathbf{K}}_{k+1} = \boldsymbol{\mathbf{P}}_{k+1}^{x,z}\boldsymbol{\mathbf{S}}_{k+1}^{-1}
        \end{equation}    
        Using the above Kalman Gain, and observation $z_{k+1}$ the updated mean is:
        \begin{equation}\label{x_meas_eq}
            \centering
            \hat{\boldsymbol{x}}_{{k+1}{|}{k+1}}=\hat{\boldsymbol{x}}_{{k}{|}{k+1}} + \boldsymbol{\mathbf{K}}_{k+1}(z_{k+1}-\hat{\boldsymbol{z}}_{{k+1}{|}{k}})
        \end{equation}
        Finally, the  covariance matrix is updated using the Kalman gain~\eqref{k_eq}:
        \begin{equation}\label{P_mean_eq}
            \centering
            \boldsymbol{\mathbf{P}}_{{k+1}{|}{k+1}} = \boldsymbol{\mathbf{P}}_{{k+1}{|}{k}}-\boldsymbol{\mathbf{K}}_{k+1}\boldsymbol{\mathbf{S}}_{k+1}{\boldsymbol{\mathbf{K}}_{k+1}}^{T} 
        \end{equation}
\end{enumerate}

\subsection{The UKF accuracy}\label{ukf_accuracy_alg}
The UKF mean estimation accuracy at the time update phase is up to the second order and in the case of Gaussian distribution, up to third order.\\
The mean zero order is matched due to the following property:
\begin{equation}\label{sp_w_1st_prop_eq}
    \centering
    \sum_{i=0}^{2n} w_{i}^{m} = 1
\end{equation}
Considering the sigma points generation \eqref{sp_0_eq}-\eqref{sp_n_i_eq}, we denote the deviations of the sigma point $i$ from the estimated mean by $\boldsymbol\delta_i$. Let 
$\mathbf{\boldsymbol\delta}_0 = 0$ be the deviation for $\boldsymbol{x}_0$, thus, for $\boldsymbol{x}_i$, $i=1,...,2n$:
\begin{equation}\label{sp_deviation_1_eq}
    \centering
    \mathbf{\boldsymbol\delta}_i = \left(\sqrt{({n}+\lambda)\boldsymbol{\mathbf{P}}_{{k}{|}{k}}}\right)_i, i = 1,...,n    
\end{equation}
\begin{equation}\label{sp_deviation_2_eq}
    \centering
    \mathbf{\boldsymbol\delta}_i = -\left(\sqrt{({n}+\lambda)\boldsymbol{\mathbf{P}}_{{k}{|}{k}}}\right)_{i-n}, i = n+1,...,2n
\end{equation}
The mean first order is matched (and odd orders are matched in case of Gaussian distribution) due to the following property:
\begin{equation}\label{sp_2nd_prop_eq}
    \centering
    \sum_{i=0}^{2n} w_{i}^{m}\mathbf{\boldsymbol\delta}_i = 0,
\end{equation}
and the mean second order is matched, due to the following property:
\begin{equation}\label{sp_p_3rd_prop_eq}
    \centering
    \sum_{i=0}^{2n} w_{i}^{m}\mathbf{\boldsymbol\delta}_i{\mathbf{\boldsymbol\delta}_i}^T = \mathbf{\boldsymbol{P}_{{k}{|}{k}}}
\end{equation}
When expressing the covariance using the Taylor expansion, the above properties yield a covariance accuracy estimation of up to the second order and in the case of Gaussian distribution, up to the third order. To compensate on the lack of higher order information 
\cite{scaledUT2002}(pending on the nonlinear dynamic system), $\gamma$ \eqref{w_c_0_eq} is added to the zero weight for the covariance calculation.
Due to the symmetric placement and equal weighting of the sigma point pairs ($\boldsymbol{x}_{i,{{k}{|}{k}}}, \boldsymbol{x}_{n+i,{{k}{|}{k}}}$), all odd orders contributions vanish and the scaling factors cancel from the second order terms. Consequently, the first scaling dependent terms, in both the mean \eqref{tu_x_mean_eq} and the time update covariance matrix \eqref{tu_p_eq}, occur at fourth order \cite{scaledUT2002}. It allows the scaling parameters to be adjusted without modifying the first and second order information represented by the sigma point set. The scaling parameters can be tuned to improve the representation of the higher order behavior of the underlying nonlinear transformation, changing the magnitude of the deviation of each sigma point from the mean, along the associated column of the resulting square root covariance matrix as constructed in \eqref{sp_i_eq} and \eqref{sp_n_i_eq}.

% ------------------------ Proposed Approach
\section{Proposed Approach}\label{prop_appr}
\begin{figure}[h]
    \centering
    \includegraphics[width=7cm, height=11cm]{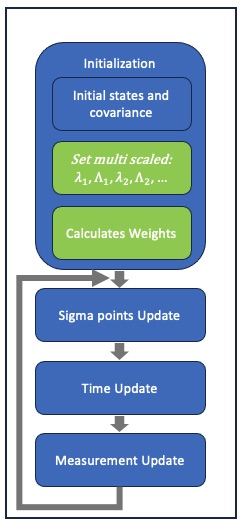}
    \caption{The MS-UKF flowchart. The new MS parameters within the initialization phase are highlighted in green, while the main UKF cycle remains the same.}
    \label{fig:MS-UKF-Flow}
\end{figure}

The UKF uses the sigma points to capture the mean and the covariance of random variables propagated through a nonlinear system up to the second order and for Gaussian random variables, up to the third order \cite{WanUkfnonlin2000}. The spreading of the sigma points is crucial for the filter to capture the actual behavior of the states. To this end, the UKF controls the spreading through a set of scaling parameters that are equal to all states. This scaling preserves the UKF accuracy and has no effect on the computation complexity. However, when states propagate through different nonlinear functions, the current formulation requires a uniform scaling set for all states, a constraint that highlights a limitation of the existing UKF formulation. To address this, we propose an extended scaling set that can be independently assigned to each state, maintaining theoretical accuracy with negligible additional computational cost. Our proposed MS-UKF flow is described in Figure \ref{fig:MS-UKF-Flow}, with our additional blocks to the standard UKF colored in green. 
To apply multi-scaling, we rewrite \eqref{lambda_std_eq} such that instead of $\alpha$ and $\kappa$ we now have $\alpha_i$, $\kappa_i$, $\Lambda_i$ for each state $i$:
\begin{equation}\label{lambda_nd_eq}
    \centering
    \lambda_i=\alpha_i^2(n+\kappa_i)-n, i=1,...,n 
\end{equation}
\begin{equation}\label{Lambda_nd_eq}
    \centering
    \Lambda_i=n + \lambda_i = \alpha_i^2(n+\kappa_i), i=1,...,n 
\end{equation}
where $\alpha_i$ determines the spread of sigma points $i$ and $n+i$ around $\hat{\boldsymbol{x}}_{{k+1}{|}{k}}$. $\kappa_i$ is a secondary scaling parameter and $\beta$ is used to incorporate prior knowledge of the distribution of $x$. Furthermore, similar to the role of \eqref{gamma_std_eq}, we define:
\begin{equation}\label{gamma_nd_eq}
    \centering
    \gamma = 1 - (\sqrt[n]{\alpha_1} \cdots \sqrt[n]{\alpha_n})^2 + \beta
\end{equation}
Equations \eqref{lambda_nd_eq}-\eqref{gamma_nd_eq} represent the UKF with multiple scaling. Notice, that small $\alpha$ implies that the sigma points might be too close to the mean, as a result $\gamma$ can increase higher order compensation and vice versa (similar to the behavior of $\gamma$ in the standard UKF), which motivates the form of \eqref{gamma_nd_eq}. Note that if $\alpha_i = \alpha$ and $\kappa_i = \kappa$ for $i=1,...,n$ (or if $n=1$), the algorithm reduces to the standard UKF. Formally, $w^c_0$ equals $w^m_0$ plus $\gamma$, so when calculating the predicted covariance, the addition of $\gamma$ results with the following covariance term:
\begin{equation}\label{p_cov_with_new_gamma_contrib_eq}
    \centering
    \begin{split}
    & \boldsymbol{\mathbf{P}}_{{k+1}{|}{k}}=
    \sum_{i=0}^{2n}w_{i}^{m}{\delta\boldsymbol{x}_{i,{{k+1}{|}{k}}} \cdot {\delta\boldsymbol{x}_{i,{{k+1}{|}{k}}}}^T} +\\  &\underline{\gamma[\delta\boldsymbol{x}_{0,{{k+1}{|}{k}}} \cdot {\delta\boldsymbol{x}_{0,{{k+1}{|}{k}}}}^T]} + \boldsymbol{\mathbf{Q}}_{k+1} 
    \end{split}
\end{equation}
The weights for the mean and the covariance \eqref{w_m_0_eq}-\eqref{w_i_eq} are extended to the MS-UKF:
 \begin{equation}\label{w_i_nd_eq}
    \centering
    w_{i}^{m} = w_{i}^{c} = \frac{1}{2\Lambda_i}, {i=1,...,n}
\end{equation}
 \begin{equation}\label{w_n_i_nd_eq}
    \centering
    w_{n+i}^{m} = w_{n+i}^{c} = \frac{1}{2\Lambda_i}, {i=1,...,n}
\end{equation}
\begin{equation}\label{w_m_0_nd_eq}
    \centering
    w_{0}^{m} = 1- \sum_{i=1}^{2n}\frac{1}{(2\Lambda_i)}
\end{equation}
\begin{equation}\label{w_c_0_nd_eq}
    \centering
    w_{0}^{c} = w_{0}^{m} + \gamma
\end{equation}
Note that \eqref{w_n_i_nd_eq} sets weights $i=n+1,...,2n$ to be the same as the weights from $i=1,...,n$ accordingly, for symmetry. Since the weights $w_i^m, i=1,...,2n$ must be positive, the following is required:
\begin{equation}\label{w_i_positive_cond_nd_eq}
    \centering
    \begin{split}
    & \alpha_i > 0, i=1,...,n\\
    & \kappa_i < n, i=1,...,n
    \end{split}
\end{equation}
The set of equations \eqref{lambda_nd_eq}-\eqref{w_c_0_nd_eq} describe a new set of scaling parameters and the appropriate weights, allowing to define one scaling set for each state, enabling to better capture the estimated mean and the covariance, while considering that some states may be propagated through different nonlinear functions and are the core of our proposed MS-UKF.\\
The MS-UKF sigma points are generated as follows:
\begin{equation}\label{sp_0_n_d_meas_nd_eq}
    \centering
    \boldsymbol{x}_{0,{{k}{|}{k}}} = \hat{\boldsymbol{x}}_{{k}{|}{k}}
\end{equation}
\begin{equation}\label{sp_i_n_d_meas_nd_eq}
    \centering
    \boldsymbol{x}_{i,{{k}{|}{k}}} = \hat{\boldsymbol{x}}_{{k}{|}{k}} + \mathbf{\boldsymbol{\delta}}_i, {{i}={1},...,{n}}
\end{equation}
\begin{equation}\label{sp_n_i_meas_nd_eq}
    \centering
    \boldsymbol{x}_{i,{{k}{|}{k}}} = \hat{\boldsymbol{x}}_{{k}{|}{k}} - \mathbf{\boldsymbol{\delta}}_i, {{i}={n+1},...,{2n}}
\end{equation}
where the deviations ($\boldsymbol{\delta}_i$) of sigma point $i$ from the estimated mean are:\\
\begin{equation}\label{sp_deviation_0_nd_eq}
    \centering
    \mathbf{\boldsymbol\delta}_0 = 0    
\end{equation}
\begin{equation}\label{sp_deviation_1_nd_eq}
    \centering
    \mathbf{\boldsymbol\delta}_i = \left(\sqrt{\Lambda_i\boldsymbol{\mathbf{P}}_{{k}{|}{k}}}\right)_i, i = 1,...,n
\end{equation}
\begin{equation}\label{sp_deviation_2_nd_eq}
    \centering
    \mathbf{\boldsymbol\delta}_i = \left(-\sqrt{\Lambda_i\boldsymbol{\mathbf{P}}_{{k}{|}{k}}}\right)_{i-n}, i = n+1,...,2n
\end{equation}
The MS-UKF sigma points generation procedure at time step $k+1$ is described in Algorithm \ref{alg:sigma points update}.
\begin{algorithm}[H]
\caption{MS-UKF sigma points generation at time step $k+1$}
\label{alg:sigma points update}
\begin{algorithmic}[1]
\State \textbf{Input:} $\boldsymbol{\mathbf{P}}_{{k}{|}{k}}, \hat{\boldsymbol{x}}_{{k}{|}{k}}, \Lambda_i, i=1,...n$.
\State \textbf{Output:} $\boldsymbol{x_{i,k|k}},i=0,...,2n$
\Statex
\State \textbf{Initialize:} $\boldsymbol{x}_{0,{{k}{|}{k}}} \gets \hat{\boldsymbol{x}}_{{k}{|}{k}}$
\For{$i = 1$ to $n$}

    \State Generate sigma point $i$:
    \[
        \boldsymbol{x}_{i,{{k}{|}{k}}} \gets \hat{\boldsymbol{x}}_{{k}{|}{k}} + \left(\sqrt{\Lambda_i\boldsymbol{\mathbf{P}}_{{k}{|}{k}}}\right)_i
    \]
    \State Generate sigma point $n+i$:
    \[
        \boldsymbol{x}_{n+i,{{k}{|}{k}}} \gets \hat{\boldsymbol{x}}_{{k}{|}{k}} - \left(\sqrt{\Lambda_i\boldsymbol{\mathbf{P}}_{{k}{|}{k}}}\right)_i
    \]
\EndFor
\end{algorithmic}
\end{algorithm}
Next, we prove that MS-UKF preserves the same properties of the UKF, as described in Section \ref{prob_form_sec}.\ref{ukf_accuracy_alg}. Namely, we shall prove that the properties \eqref{sp_w_1st_prop_eq}, \eqref{sp_2nd_prop_eq} and \eqref{sp_p_3rd_prop_eq}, considering the extended structure, are satisfied. Thus, we show that the proposed scaling set increases UKF abilities without reducing its accuracy.
\begin{theorem}\label{Theorem1}
In the MS-UKF the mean zero order is matched.
Let:
 \begin{equation}\label{theorem_1_w_i_nd_eq}
    \centering
    w_{i}^{m} = \frac{1}{2\Lambda_i}, {i=1,...,n},
\end{equation}
 \begin{equation}\label{theorem_1_w_n_i_nd_eq}
    \centering
    w_{n+i}^{m} = \frac{1}{2\Lambda_i}, {i=1,...,n},
\end{equation}
and
\begin{equation}\label{theorem_1_w_m_0_nd_eq}
    \centering
    w_{0}^{m} = 1- \sum_{i=1}^{2n}\frac{1}{(2\Lambda_i)}
\end{equation}
Than:
\begin{equation}\label{theorem_1_sum_w_1_eq}
    \centering
        \sum_{i=0}^{2n} w_{i}^{m} = 1
\end{equation}
\begin{proof}
\begin{equation}\label{sp_w_1st_prop_nd_eq}
    \begin{split}
    \centering
        & \sum_{i=0}^{2n} w_{i}^{m} = w_{0}^{m} + \sum_{i=1}^{2n}\frac{1}{(2\Lambda_i)} = \\
        & 1-\sum_{i=1}^{2n}\frac{1}{(2\Lambda_i)} + \sum_{i=1}^{2n}\frac{1}{(2\Lambda_i)} = 1
    \end{split}
\end{equation}
\end{proof}
\end{theorem}
\begin{theorem}\label{Theorem2}
In the MS-UKF the mean first order is matched.
Let $\boldsymbol\delta_i$, $i=0,...2n$ be defined as:
\begin{equation}\label{th_2_sp_deviation_0_nd_eq}
    \centering
    \begin{split}
    &\mathbf{\boldsymbol\delta}_0 = 0\\
    &\mathbf{\boldsymbol\delta}_i = \left(\sqrt{\Lambda_i\boldsymbol{\mathbf{P}}_{{k}{|}{k}}}\right)_i, i = 1,...,n\\
    &\mathbf{\boldsymbol\delta}_i = -\left(\sqrt{\Lambda_i\boldsymbol{\mathbf{P}}_{{k}{|}{k}}}\right)_{i-n}, i = n+1,...,2n
    \end{split}
\end{equation}
Than,
\begin{equation}\label{theorem_2_sum_w_delta_0_eq}
    \centering
    \sum_{i=0}^{2n} w_{i}^{m}\mathbf{\boldsymbol\delta}_i = 0 
\end{equation}
\begin{proof}
\leavevmode

From \eqref{sp_deviation_0_nd_eq}, $\boldsymbol\delta_0=0$, thus
\begin{equation}\label{sp_w_2nd_prop_nd_eq_1}
    \begin{aligned}
    \sum_{i=0}^{2n} w_{i}^{m}\mathbf{\boldsymbol\delta}_i =\sum_{i=1}^{2n} w_{i}^{m}\mathbf{\boldsymbol\delta}_i
    \end{aligned}
\end{equation}
Substituting \eqref{w_i_nd_eq} and \eqref{w_n_i_nd_eq} into \eqref{sp_w_2nd_prop_nd_eq_1}, we obtain:\\
\begin{equation}\label{sp_w_2nd_prop_nd_eq_2}
    \centering
    \begin{aligned}
       \sum_{i=1}^{2n} w_{i}^{m}\mathbf{\boldsymbol\delta}_i=&\sum_{i=1}^{n} \mathbf{\frac{1}{(2\Lambda_i)}}\mathbf{\boldsymbol\delta}_i + \sum_{i=n+1}^{2n} \mathbf{\frac{1}{(2\Lambda_i)}}\mathbf{\boldsymbol\delta}_i =\\  
       &\sum_{i=1}^{n} \mathbf{\frac{1}{(2\Lambda_i)}}\mathbf{\boldsymbol\delta}_i - \sum_{i=1}^{n} \mathbf{\frac{1}{(2\Lambda_i)}}\mathbf{\boldsymbol\delta}_i = 0 
    \end{aligned}
\end{equation}
\end{proof}
\end{theorem}

\begin{lemma}\label{lemma_1_w_delta_P}
For $i \neq 0$
    \begin{equation}\label{sp_w_delta_eq}
        \centering
        w_{i}^{m}\mathbf{\boldsymbol\delta}_i(\mathbf{\boldsymbol\delta}_i)^T = \mathbf{\frac{1}{2}}\left(\sqrt{\boldsymbol{\mathbf{P}}_{{k}{|}{k}}}\right)_i\left(\sqrt{\boldsymbol{\mathbf{P}}_{{k}{|}{k}}}\right)_i ^T
    \end{equation}
\begin{proof}
\leavevmode

Substituting \eqref{sp_deviation_1_nd_eq} and \eqref{sp_deviation_2_nd_eq} into Lemma \ref{lemma_1_w_delta_P} yields:
\begin{equation}
    \centering
    \begin{aligned}
    \quad w_{i}^{m}\mathbf{\boldsymbol\delta}_i(\mathbf{\boldsymbol\delta}_i)^T &=\mathbf{\frac{1}{2\Lambda_i}}\left(\sqrt{\Lambda_i\boldsymbol{\mathbf{P}}_{{k}{|}{k}}}\right)_i\left(\sqrt{\Lambda_i\boldsymbol{\mathbf{P}}_{{k}{|}{k}}}\right)_i ^T\\
    &=\mathbf{\frac{1}{2}}\left(\sqrt{\boldsymbol{\mathbf{P}}_{{k}{|}{k}}}\right)_i\left(\sqrt{\boldsymbol{\mathbf{P}}_{{k}{|}{k}}}\right)_i ^T
    \end{aligned}
    \end{equation}
\end{proof}
\end{lemma}
\begin{theorem}\label{Theorem3}
In the MS-UKF the mean second order is matched.\\
Let $\boldsymbol\delta_i$, $i=0,...,2n$ be defined as \eqref{sp_deviation_0_nd_eq}-\eqref{sp_deviation_2_nd_eq}, then
\begin{equation}\label{theorem_3_sum_w_delta_delta_p_eq}
    \centering
    \sum_{i=0}^{2n} w_{i}^{m}\mathbf{\boldsymbol\delta}_i(\mathbf{\boldsymbol\delta}_i)^T = \boldsymbol{\mathbf{P}}_{{k}{|}{k}} 
\end{equation}
\begin{proof}
\leavevmode

Since $\boldsymbol\delta_0=0$:
    \begin{equation}\label{sp_p_3nd_prop_nd_eq_1}
        \centering
        \begin{aligned}
           &\sum_{i=0}^{2n} w_{i}^{m}\mathbf{\boldsymbol\delta}_i(\mathbf{\boldsymbol\delta}_i)^T =\sum_{i=1}^{n} w_{i}^{m}\mathbf{\boldsymbol\delta}_i(\mathbf{\boldsymbol\delta}_i)^T + \sum_{i=n+1}^{2n} w_{i}^{m}\mathbf{\boldsymbol\delta}_i(\mathbf{\boldsymbol\delta}_i)^T
        \end{aligned}
    \end{equation}
Using Lemma \ref{lemma_1_w_delta_P} yields:
    \begin{equation}\label{sp_p_3nd_prop_nd_eq_2}
        \centering
        \begin{aligned}
           &\sum_{i=1}^{n} w_{i}^{m}\mathbf{\boldsymbol\delta}_i(\mathbf{\boldsymbol\delta}_i)^T + \sum_{i=n+1}^{2n} w_{i}^{m}\mathbf{\boldsymbol\delta}_i(\mathbf{\boldsymbol\delta}_i)^T =\\
           &\mathbf{\frac{1}{2}}\sum_{i=1}^{n}\left(\sqrt{\boldsymbol{\mathbf{P}}_{{k}{|}{k}}}\right)_i\left(\sqrt{\boldsymbol{\mathbf{P}}_{{k}{|}{k}}}\right)_i ^T +\\
           &\mathbf{\frac{1}{2}}\sum_{i=n+1}^{2n}\left(\sqrt{\boldsymbol{\mathbf{P}}_{{k}{|}{k}}}\right)_i\left(\sqrt{\boldsymbol{\mathbf{P}}_{{k}{|}{k}}}\right)_i ^T\\
        \end{aligned}
    \end{equation}
Using the Cholesky decomposition, we can write:
    \begin{equation}\label{sp_p_3nd_prop_nd_eq_3}
        \centering
        \begin{aligned}
           &\mathbf{\frac{1}{2}}\sum_{i=1}^{n}\left(\sqrt{\boldsymbol{\mathbf{P}}_{{k}{|}{k}}}\right)_i\left(\sqrt{\boldsymbol{\mathbf{P}}_{{k}{|}{k}}}\right)_i ^T +\\
           &\mathbf{\frac{1}{2}}\sum_{i=n+1}^{2n}\left(\sqrt{\boldsymbol{\mathbf{P}}_{{k}{|}{k}}}\right)_i\left(\sqrt{\boldsymbol{\mathbf{P}}_{{k}{|}{k}}}\right)_i ^T=\\
           &\mathbf{\frac{1}{2}}\boldsymbol{\mathbf{P}}_{{k}{|}{k}}+\mathbf{\frac{1}{2}}\boldsymbol{\mathbf{P}}_{{k}{|}{k}} = \boldsymbol{\mathbf{P}}_{{k}{|}{k}} 
        \end{aligned}
    \end{equation}
\end{proof}
\end{theorem}
Assuming the deviations \eqref{sp_deviation_1_nd_eq} and \eqref{sp_deviation_2_nd_eq} are zero mean Gaussian, satisfying those properties, enables the MS-UKF to estimate the mean up to the 3rd order, as shown in Appendix \ref{appendix_A_sec}. It also guaranties that the unbiased accuracy remains the same as the standard UKF. Further, as in the standard UKF, the MS-UKF covariance estimation is up to the 3rd order as shown in Appendix \ref{appendix_B_sec}.
Similar to the UKF, due to the symmetric placement and equal weighting of the sigma point pairs ($\boldsymbol{x}_{i,{{k}{|}{k}}}, \boldsymbol{x}_{n+i,{{k}{|}{k}}}$), all odd orders contributions vanish and the scaling factors cancel from the second order terms. Consequently, the first scaling dependent terms in both the mean \eqref{tu_x_mean_eq} and the time update covariance matrix \eqref{p_cov_with_new_gamma_contrib_eq} occur at fourth order. It allows the scaling parameters to be adjusted without modifying the first and second order information represented by the sigma-point set. The scaling parameters can be tuned to improve the representation of the higher order behavior of the underlying nonlinear transformation, changing the magnitude of the deviation of each sigma point pair from the mean differently, along the associated column of the resulting square root covariance matrix as constructed in \eqref{sp_deviation_1_nd_eq} and \eqref{sp_deviation_2_nd_eq}.
The proposed scaling method requires only a small set of additional computations during the UKF initialization phase as can be seen in Figure \ref{fig:MS-UKF-Flow}, where the affected steps are in green within the initialization phase. It requires very small additional resource keeping the different weights $w^{m/c}_i=1/{2\Lambda_i}, i=1,...,n$  where in the standard UKF they are all the same ($w^{m/c}_i=1/{2\Lambda}, i=1,...,n$). No additional computations are needed during the rest of the MS-UKF life cycles.\\
The MS-UKF scaling system extends possibilities to better fit the underlying model. Though, similar to the UKF, not all values are valid. We already introduced in \eqref{w_i_positive_cond_nd_eq} the conditions for the weights $w_i^c,w_i^m, i=1,...,n$ to be positive. In addition, from \eqref{p_cov_with_new_gamma_contrib_eq}, a sufficient condition for the MS-UKF propagated covariance matrix $\boldsymbol{\mathbf{P}}_{{k+1}{|}{k}} \in \mathbb{R}^{n \times n}$ at time step $k+1$ ($k \geq 0$) to be positive semi-definite is:
\begin{equation}\label{scaling_system_psd_criteria_eq}
    \centering
    \left( \sum_{i=1}^{n}\frac{1}{\alpha^2_i (n+\kappa_i)} \right)^{-1}-(\sqrt[n]{\alpha_1} \cdots \sqrt[n]{\alpha_n})^2 + \beta \geq 0
\end{equation}
where $n > 0$ is the size of the state vector. For a detailed description, please refer to Appendix \ref{appendix_C_sec}.\\
If $\alpha_i = \alpha$ and $\kappa_i=\kappa$, $i=1,...,n$ we get from \eqref{scaling_system_psd_criteria_eq}:
\begin{equation}\label{scaling_system_psd_criteria_for_ukf_eq}
    \centering
    \begin{split}
    & \left( \sum_{i=1}^{n}\frac{1}{\alpha^2 (n+\kappa)} \right)^{-1}-\alpha^2 + \beta =  \\
    & \frac{\alpha^2 (n+\kappa)}{n}-\alpha^2 + \beta =  \\
    & \frac{\alpha^2\kappa}{n} + \beta \geq 0
    \end{split}
\end{equation}
which is the UKF scaling limitation.\\
For the UKF, the paper \cite{WanUkfnonlin2000} suggests $\alpha$ to be given a small positive value ($1e-3$), $\kappa$ is usually set to $0$, and $\beta$ is used to incorporate prior knowledge of the distribution of the state vector $x$ ($\beta=2$ for Gaussian distribution).\\
For the MS-UKF, the same principles hold, meaning $\alpha_i \ll 1$, $\kappa_i=0$ for $i=1,...,n$ and $\beta$ depends on the distribution.\\
Note that if $\kappa_i=0$ for $i=1,...,n$ and $\beta=2$, from \eqref{scaling_system_psd_criteria_eq} we get:
\begin{equation}\label{scaling_system_psd_criteria_kappa_i_0_eq}
    \centering
    \frac{n}{\sum_{i=1}^{n}\frac{1}{\alpha^2_i}}-(\alpha_1^2 \cdots \alpha_n^2)^\frac{1}{n} + 2 \geq 0
\end{equation}
The left term is the harmonic mean of $\alpha_i^2$ and the middle term is the geometric mean of $\alpha_i^2$, in that case, for $\alpha_i \leq 1, i=1,...,n$ the condition is met, meaning the criteria is reasonable considering common scaling usages, other options should be considered using \eqref{scaling_system_psd_criteria_eq}.

% ------------------------ Results
\section{Analysis and Results}\label{res_sec}
To show the advantages of our approach we selected two multi-state nonlinear examples.
\subsection{Performance Metrics}\label{results_perform_metrics}
We employ three performance metrics to evaluate our proposed approach:
\begin{enumerate}
    \item Root mean square error (RMSE).
    \item Total RMSE (TRMSE).
    \item Total STD errors (TSTD).
\end{enumerate}
The RMSE of state $l$ is defined by:
\begin{equation}\label{rmse_l_calc}
    \centering
    \textbf{RMSE} = \sqrt{\frac{1}{n \cdot m}\sum_{i=1}^{n}\sum_{j=1}^{m}\boldsymbol{x}_{e_{l,i,j}}^2}
\end{equation}
where $\boldsymbol{x}_{e_{l,i,j}}$ is the error of state $l$, $l=1,2$, at time step $i$ and Monte-Carlo (MC) run $j$, measuring the RMSE of each state along the experiment procedure, $m$ is the number of MC runs, and $n$ is the number of time steps.\\ 
The TRMSE is defined by:
\begin{equation}\label{total_rmse_l_calc}
    \centering
    \textbf{TRMSE} = \sqrt{{RMES_1}^2 + {RMES_2}^2}
\end{equation}
The TSTD is the root sum square of the state error STDs, calculated across $m$ Monte Carlo runs for a specific time step $n$:
\begin{equation}\label{trmse_n_calc}
    \centering
    \textbf{TSTD} = \sqrt{\frac{1}{m}\sum_{j=1}^{m}(\boldsymbol{x}_{e_{1,n,j}}-\mu_{e_{1,n}})^2+(\boldsymbol{x}_{e_{2,n,j}}-\mu_{e_{2,n}})^2}
\end{equation}
where $\mu_{e_{l,n}}$ is the mean error of state $l$ across MC runs at time step $n$.
% Example 1
\subsection{Example I}\label{results_example_1}
The first model is a two dimensional sigmoid system, dynamically independent but coupled through the measurement model. They share similar functionalities but different uncertainties, one should be captured using a relative large spreading of the appropriate sigma points and the other using a small spreading. Since the standard UKF is limited to a single scaling set, it must compromise, while the MS-UKF can optimize the spreading differently for each state. Such model commonly appears in practical issues such as trying to estimate the input inner states out of the resulting activations as dynamic causal modeling \cite{dcm_2003}.\\

\subsubsection{System model and measurement}\label{res_ex1_model_and_meas}
\begin{figure}[h]
    \centering
    \centering
    \includegraphics[width=8.5cm, height=8cm]{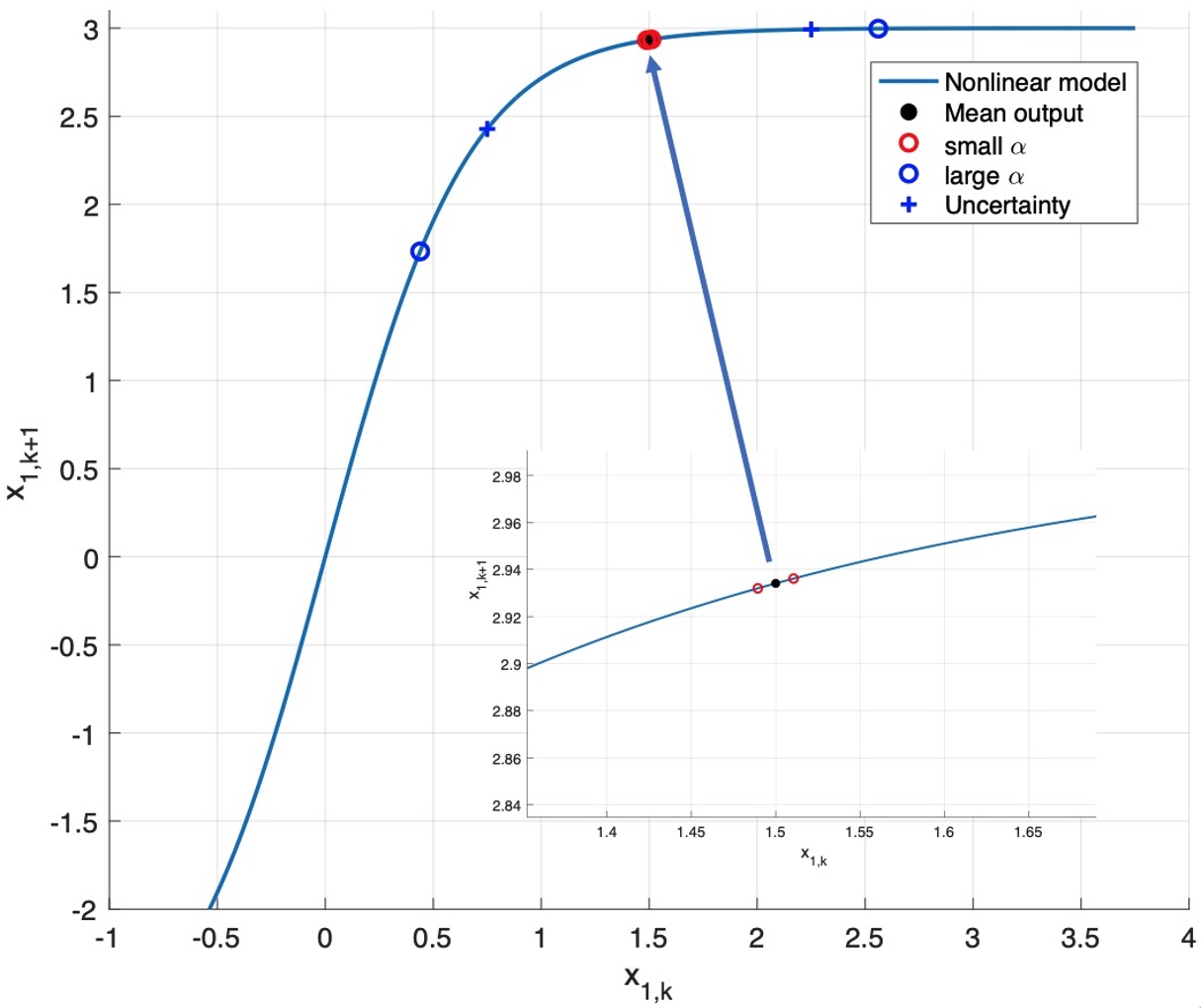}
    \caption{Sigma points extraction for $x_1$ as function of $\alpha$ (0.01 vs 1.0) - larger $\alpha$ spreads fits the model better.}
    \label{fig:sp_extract_around_x1}
\end{figure}

\begin{figure}[h]
    \centering
    \centering
    \includegraphics[width=8.5cm, height=8cm]{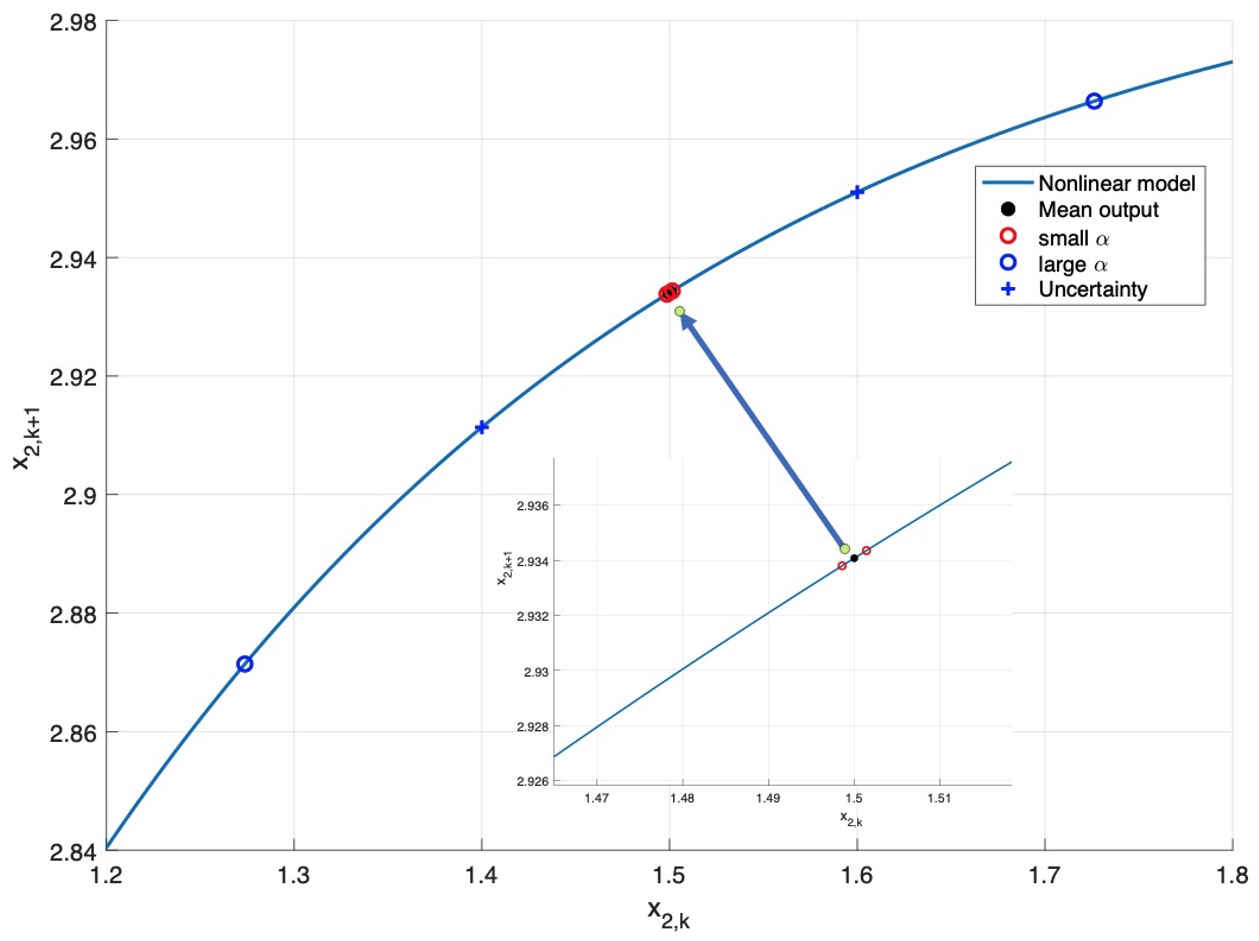}
    \caption{Sigma points extraction for $x_2$ as function of $\alpha$ (0.01 vs 1.0) - smaller $\alpha$ spreads fits the model better.}
    \label{fig:sp_extract_around_x2}
\end{figure}

The system model is as follows:
\begin{equation}\label{2d_sig_pred_model_eq}
    \centering
    \mathbf{x}_{i,k+1} = a_i\cdot\Delta t \cdot sig(g\mathbf{x}_{i,k}) + b_i + {\omega_k}_i,  i=1,2
\end{equation}
where $k$ is the time step, $\mathbf{x_k} \in \mathbb{R} ^2$ is the state vector, $a_i, b_i \in \mathbb{R}, \omega_k\sim\mathcal{N}(0, \mathbf{Q})$ is a zero-mean white Gaussian noise with covariance,
\begin{equation}\label{ex1_q_meas_eq}
    \centering
    \mathbf{Q} = 
    \begin{bmatrix}
    q_1&0\\
    0&q_2\\
    \end{bmatrix},
\end{equation}
and
\begin{equation}\label{sig_eq}
    \centering
    sig(\mathbf{x}) = \frac{1}{1+e^{-\mathbf{x}}}
\end{equation}
The measurement model is:
\begin{equation}\label{2d_sig_meas_model_ex1_eq}
    \centering
    \mathbf{z}_{k+1} =  \mathbf{H} \cdot \mathbf{x}_k + v_k 
\end{equation}
where
$v_k \sim\mathcal{N}(0, \mathbf{R})$ is a zero-mean white Gaussian noise with covariance,
\begin{equation}\label{ex1_r_meas_eq}
    \centering
    \mathbf{R} = 
    \begin{bmatrix}
    r_1&0\\
    0&r_2\\
    \end{bmatrix},
\end{equation}
and the measurement matrix is:
\begin{equation}\label{ex1_h_meas_eq}
    \centering
    \mathbf{H} = 
    \begin{bmatrix}
    1&0.1\\
    0.1&1\\
    \end{bmatrix},
\end{equation}
The initial conditions are:
\begin{equation}\label{ex1_settings_eq}
    \centering
    \begin{split}
    & \Delta t=0.05 [sec],\\
    & duration=600 [\text{timesteps}],\\
    & X_0=[1.5,1.5]^T,\\
    & a=[120.0, 120.0],\\
    & b=[-3.0, -3.0],\\
    & r_1=0.75^2, r_2=0.15^2,\\
    & g=3.0,\\
    & q_1=0.5, q_2=0.05\\
    & \boldsymbol{\mathbf{P}}_{0|0} = 
    \begin{bmatrix}
    2.5&0\\
    0&0.1\\
    \end{bmatrix}
    \end{split}
\end{equation}
where $\Delta t$ is the time step in seconds, $duration$ is the duration in time steps, and $\boldsymbol{\mathbf{P}}_{0|0}$ reflects different uncertainty for each state. Note that due to the initial covariance, the behavior of the model around $x_{1,0}$ is different from the behavior around $x_{2,0}$. Figures \ref{fig:sp_extract_around_x1} and \ref{fig:sp_extract_around_x2} illustrate the effect of $\alpha$ on the extraction of the sigma points and the impact due to the nonlinearity models.
In the case of the first state, with large $\alpha$ (1.0) the sigma points fits the model better compared to the small $\alpha$ (0.01), so the propagated mean estimation will be closer to the real one (Figure \ref{fig:sp_extract_around_x1}). In the second state, with small $\alpha$ (0.01) the sigma points fits the model better than a large $\alpha$ (1.0), as shown in Figure \ref{fig:sp_extract_around_x2}. Thus, when we are limited to a single scaled factor, compromise is needed. With the MS-UKF we can assign a large $\alpha$ to the first state and a smaller $\alpha$ to the second, achieving better performance.\\
\\

\subsubsection{Results}\label{res_ex1_res}
\begin{figure}[h]
    \centering
    \centering
    \includegraphics[width=9cm, height=8cm]{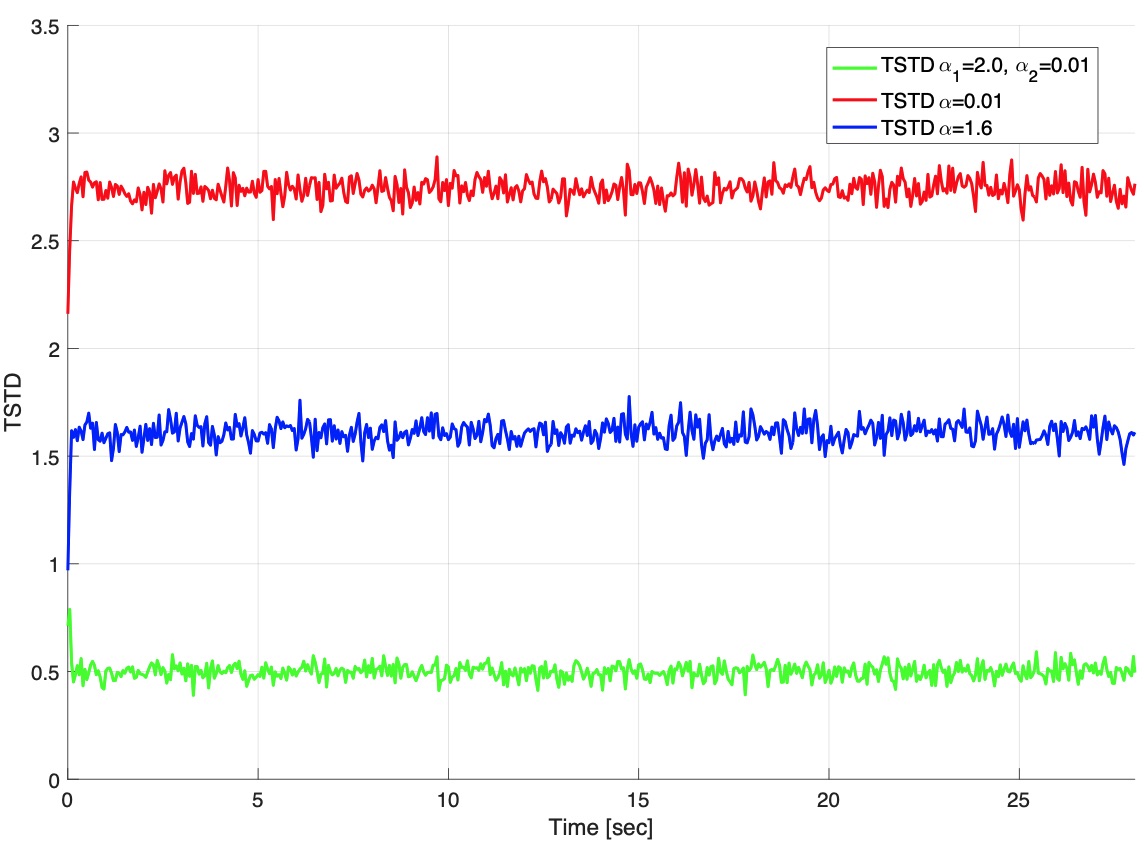}
    \caption{TSTD comparison between two UKFs with different $\alpha$ values (red and blue) and our MS-UKF (green).}
    \label{fig:comparison_of_normalized_error_STD_across_MC_runs}
\end{figure}

We have concentrated on $\alpha$ since its impact is on both the mean and the covariance estimation (fixing $\beta=2.0$ and $\kappa=0$). In the standard UKF, we are limited to a single $\alpha$. Thus, we examined values from $0.01$ to $2.0$ in increments of $0.1$. For each value, we performed 100 MC runs, compared the performance (minimum RMSE) across all states, and selected the $\alpha$ that yielded the greatest improvement. The motivation for the criteria is due to the fact that in this model, the states may not present the same element. Consequently, the contribution to the total behavior may be different. Thus, we have calculated the RMSE of each state, than calculated the improvement factor for each state and continued with $\alpha$ that achieved the largest improvement. The search ended with $\alpha=1.6$ as the optimal value. To find the optimal valid combination of $\alpha_1$ and $\alpha_2$, we have executed a search for both $(\alpha_1, \alpha_2)$, keeping the same search increment. The two dimensional search results ended up with $\alpha_1=2.0$ and $\alpha_2=0.01$, as the optimal values. For comparison, we have executed MC runs also for the case where $\alpha=0.01$. Figure \ref{fig:comparison_of_normalized_error_STD_across_MC_runs} shows the TSTD per time step across MC runs along the procedure for each of the selected scaling. Table \ref{tab:sig_comp_tbl} summarizes the results, showing that the MS-UKF achieved an improvement of $81.8\%$ in the case of $\alpha=0.01$ and $68.8\%$ in the case of $\alpha=1.6$. The rest of the tuning remains the same to emphasize the importance of being able to better cope the model, using the enhanced scaling approach.\\

\begin{table}[h]
    \caption{TSTD the MS-UKF, UKF with $\alpha=0.01$, and UKF with $\alpha=1.6$ applied to the 2D sigmoid model.}
    \centering
    \begin{tabular}{|p{4.5cm}|p{1.0 cm}|p{1.4cm}|}
    \hline
    Method & TSTD errors & MS-UKF improvement\\
    \hline
    \textbf{MS-UKF(ours) $\alpha_1=2.0,\alpha_2=0.01$} & 0.50 & - \\
    \textbf{UKF $\alpha=0.01$} & 2.74 &  81.8\% \\
    \textbf{UKF(optimal) $\alpha=1.6$} & 1.61 &   68.8\% \\
    \hline
    \end{tabular}
    \label{tab:sig_comp_tbl}
\end{table}

% Example 2
\subsection{Example II}\label{results_example_2}
The second example introduces a two-dimensional dynamic system model, where the two states are dynamically independent but coupled throughout the process by evolving covariance matrices. This model is evident in a two-axis servo system with cogging torque, such as an azimuth–elevation antenna or a pan–tilt platform.\\
\subsubsection{System model and measurement}
The system model is as follows:
\begin{equation}\label{2d_trig_pred_model_eq}
    \begin{aligned}
    & \mathbf{x}_{1,k+1} = \mathbf{x}_{1,k} + \Delta t \cdot a_1 \cdot sin(b_1 \cdot \mathbf{x}_{1,k}) + \\
    &\quad\quad\quad\quad \Delta t \cdot 0.3 \cdot sin(2 \cdot \mathbf{x}_{1,k}) + \omega_{1,k} \\
    & \mathbf{x}_{2,k+1} = \mathbf{x}_{2,k} + \Delta t \cdot a_2 \cdot cos(b_2 \cdot \mathbf{x}_{1,k}) + \omega_{2,k},
    \end{aligned}
\end{equation}
where $k$ is the time step, $\Delta t$ is the time step duration (seconds), $\mathbf{x_k} \in \mathbb{R} ^2$ is the state vector, $a_i, b_i \in \mathbb{R}, \omega_{i,k}\sim\mathcal{N}(0, q_i)$, and\\
the measurement model is:
\begin{equation}\label{2d_sig_meas_model_ex2_eq}
    \centering
    \mathbf{z}_{k+1} =  \mathbf{H} \cdot \mathbf{x}_k + v_k 
\end{equation}
where $v_k \sim\mathcal{N}(0, \mathbf{R})$ a zero-mean white Gaussian noise
with covariance, 
\begin{equation}\label{ex2_r_meas_eq}
    \centering
    \mathbf{R} = 
    \begin{bmatrix}
    r_1&0\\
    0&r_2\\
    \end{bmatrix},
\end{equation}
and the measurement matrix is:
\begin{equation}\label{ex2_h_meas_eq}
    \centering
    \mathbf{H} = 
    \begin{bmatrix}
    1&0\\
    0&1\\
    \end{bmatrix},
\end{equation}
The following setting reflects different accuracy for each axis, introducing one with additional disturbances.\\
The initial conditions are:
\begin{equation}\label{ex2_settings_eq}
    \centering
    \begin{split}
    & \Delta t=0.01,\\
    & duration=600 [\text{time steps}],\\
    & \mathbf{X}_0=[0.0, 0.0],\\
    & a=[3.0, 5.0],\\
    & b=[2.3, 3.0],\\
    & r_1=1.5^2, r_2=1.5^2,\\
    & q_1=0.001, q_2=0.01,\\
    & \boldsymbol{\mathbf{P}}_{0|0} = 
    \begin{bmatrix}
    0.7&0\\
    0&1.0\\
    \end{bmatrix}
    \end{split}
\end{equation}

\subsubsection{Results}
\begin{figure}[h]
    \centering
    \centering
    \includegraphics[width=9cm, height=7cm]{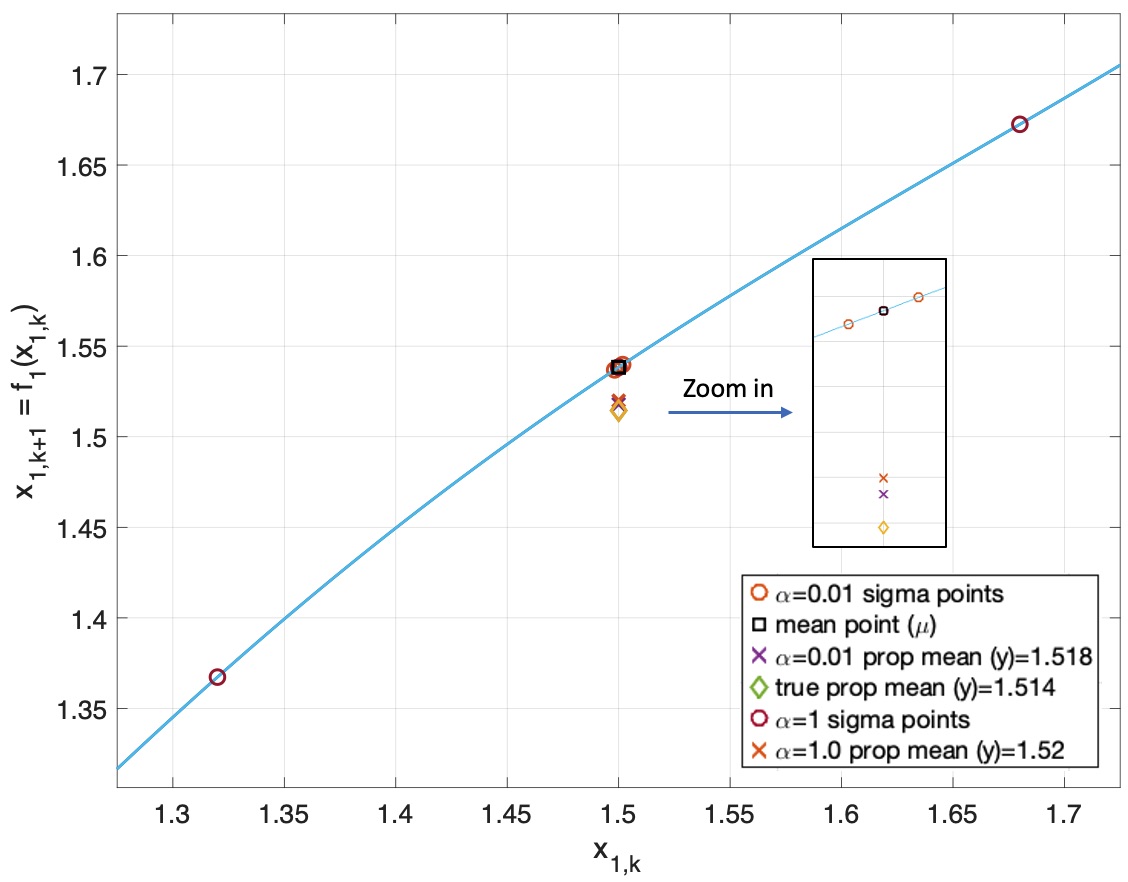}
    \caption{Sigma points estimation due to $\alpha=0.01$ and $\alpha=1.0$ after one cycle of example 2 $x_1$ axis}
    \label{fig:mean_sp_due_to_alpha_ex2}
\end{figure}

\begin{figure}[h]
    \centering
    \centering
    \includegraphics[width=9cm, height=7cm]{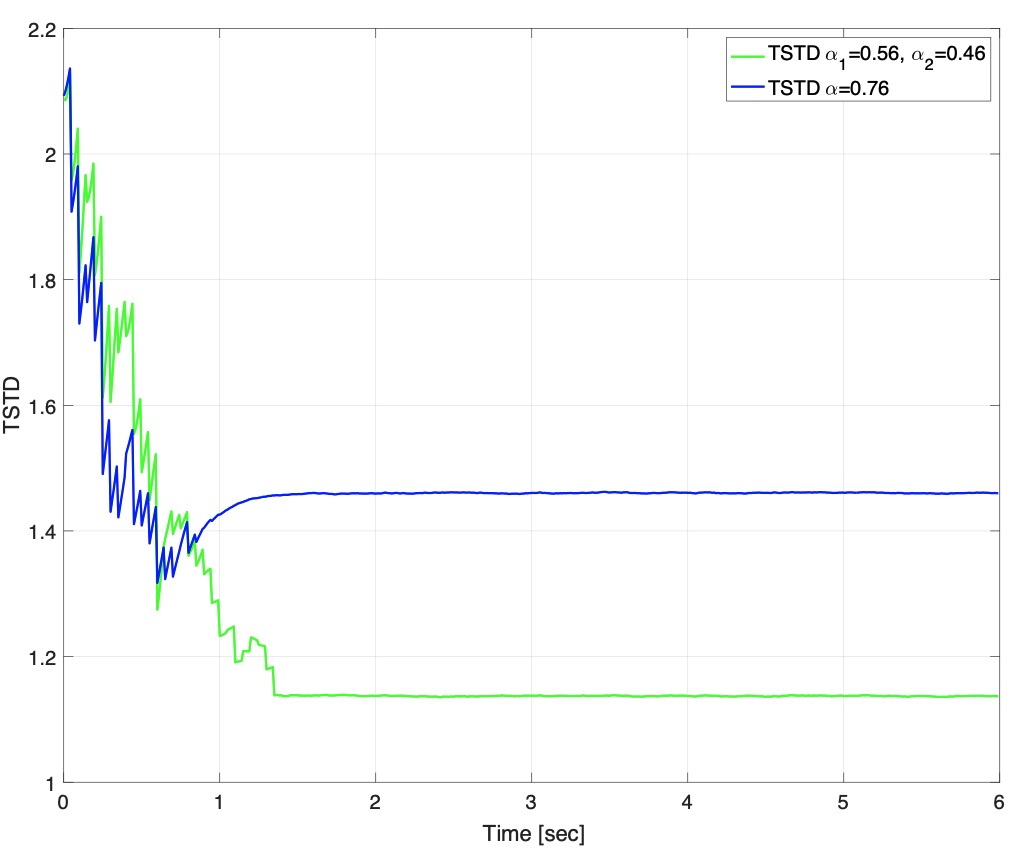}
    \caption{TSTD comparison between a UKF with a single $\alpha$ value (blue) and our MS-UKF (green).}
    \label{fig:Comparison_of_total_errors_STD_across_MC_runs_ex2}
\end{figure}

Since in this case the states represent similar elements (dynamic movement), the contribution to the overall performance of each state has similar weight. We examined values from 0.01 to 2.0 in increments of 0.05. For each value, we performed 100 MC runs, compared the performance (minimum RMSE) across both states, and selected the $\alpha$ that yielded the greatest improvement. We have calculated the $RMSE_l$ of each state \eqref{rmse_l_calc} and the TRMSE \ref{total_rmse_l_calc}. Figure \ref{fig:mean_sp_due_to_alpha_ex2} shows the differentiation in the resulting sigma points mean value, depending on the selected $\alpha$. Figure \ref{fig:Comparison_of_total_errors_STD_across_MC_runs_ex2} shows the TSTD per time step across MC runs along the procedure for each of the selected scaling. In the case of the standard UKF, the optimal value $\alpha=0.76$ achieved a TSTD of 1.47. In the case of the MS-UKF, we performed a 2D parameter search for valid $\alpha_1, \alpha_2 \in [0.01, 2.0]$ with a step size of $0.05$. The optimal configuration ($\alpha_1 = 0.56, \alpha_2 = 0.46$) produced a TSTD of $1.21$, representing a $17\%$ improvement over the baseline.

Table \ref{tab:sincos_comp_tbl} summarizes the results, showing the MS-UKF TSTD of 1.21, with the selection of $\alpha_1=0.56$ and $\alpha_2=0.46$ and the UKF with TSTD of 1.47 with the selection of a single $\alpha=0.76$, improving it by $17\%$.\\
\begin{table}[h]
    \caption{TSTD of the MS-UKF and the UKF applied to the 2D dynamic system model.}
    \centering
    \begin{tabular}{|p{4.6cm}|p{1.4cm}|p{1.1cm}|}
    \hline
    Method & TSTD errors & MS-UKF improvement. \\
    \hline
    \textbf{MS-UKF(ours) $\alpha_1=0.56,\alpha_2=0.46$} & 1.21 & - \\
    \textbf{UKF(optimal) $\alpha=0.76$} & 1.47 &  17\% \\
    \hline
    \end{tabular}
    \label{tab:sincos_comp_tbl}
\end{table}

% ------------------------ Conclusion
\section{Conclusion}\label{conc_sec}
The ability to control the spread of the sigma points around the estimated mean in the standard UKF is crucial for accurately capturing the filter behavior and improving estimation accuracy. Since in multi-dimensional models, different states may behave substantially different, we have modified the standard UKF, introducing MS-UKF that can control the spreading differently per dimension while maintaining the key properties of the sigma points. In is regard, the filter preserves the accuracy level as the standard UKF in terms of mean and covariance estimations, while fitting better the underlying model, improving performance, accuracy and filter behavior. We provided the mathematical formulation of the MS-UKF extended structure, proving that the fundamental properties of the standard UKF in terms of estimation accuracy are preserved. It has been demonstrated that the proposed modification demands minimal additional computational overhead or resources. Additionally, we have demonstrated our approach on two different nonlinear dynamic systems. The first system has sigmoid functions in the core of the dynamic formulation, showing a performance improvement of $68.8\%$ over the standard UKF. The second system dynamic functions contain sine and cosine, showing a performance improvement of $17\%$ over the standard UKF.\\
The enhanced capability of the MS-UKF, due to extend tuning set of parameters, improves accuracy, but induces a design complexity to better suit the filter to the underlying system model.\\
In conclusion, many dynamic models are multi-dimensional, where the model exhibits non-uniform nonlinear behavior across states. Therefore, the MS-UKF offers multi-dimensional tuning system, that better captures the nonlinear behavior of the multi-dimensional states, improving estimation and accuracy.

% ------------------------ Appendix
\begin{appendices}

\section{}\label{appendix_A_sec}
In this section, we show that the MS-UKF is capable to estimate the mean up to the second order and in case of Gaussian distribution, up to the thrid order. To that end, let $\boldsymbol{x}$ be an $n$-dimensional random variable with mean $\boldsymbol{\mu}$ and covariance $\boldsymbol{P}_x$, $f: \mathbb{R}^n \rightarrow \mathbb{R}^n$ be a nonlinear transformation, $\boldsymbol{x}=\boldsymbol{\mu}+\boldsymbol{\delta}$, $\boldsymbol{x},\boldsymbol{\mu},\boldsymbol{\delta}\in\mathbb{R}^n$ and $\boldsymbol{\delta}$ is coming from zero mean random variable with $E[\boldsymbol{\delta}\boldsymbol{\delta}^T]=\boldsymbol{P}_x \in\mathbb{R}^{n\times n}$.\\
Thus, the Taylor expansion of $f$ around $\boldsymbol{\mu}$ is:
\begin{equation}\label{true_taylor_exp_eq}
    \centering
    \begin{split}
    &\boldsymbol{f}(\boldsymbol{x})=\boldsymbol{f}(\boldsymbol{\mu+\delta})=\\
    &\quad\boldsymbol{f}(\mu)+\nabla\boldsymbol{f}\delta + \frac{\nabla^2\boldsymbol{f} \delta^2}{2!}+\frac{\nabla^3\boldsymbol{f}\delta^3}{3!}+\frac{\nabla^4\boldsymbol{f}\delta^4}{4!}+\cdot\cdot\cdot
    \end{split}
\end{equation}
where the notation $\nabla^i\boldsymbol{f}\delta^i$ stands for the i-th order term in the multi-dimensional Taylor series (for more information see \cite{scaledUT2002}).
Taking expectation on \eqref{true_taylor_exp_eq} gives:
\begin{equation}\label{true_mean_eq}
    \centering
    E[\boldsymbol{f}(\boldsymbol{\mu+\delta})]= \boldsymbol{f}(\mu)+\frac{1}{2}\nabla^2\boldsymbol{f}P+\frac{1}{3!}\nabla^3\boldsymbol{f} E[\delta^3]+\cdot\cdot\cdot
\end{equation}
\begin{lemma}\label{lemma2}
Let $\boldsymbol{\delta}_i$ be a zero mean random variable as structured in \eqref{sp_deviation_1_nd_eq} and \eqref{sp_deviation_2_nd_eq}. Also, let $\boldsymbol{x}_i=f(\boldsymbol{\mu}+\boldsymbol{\delta}_i)$ be the MS-UKF propagated sigma points $i=0,...2n$\\
Thus,
\begin{equation}\label{lemma_2_msukf_x_mean_nd_eq}
    \centering
    \hat{\boldsymbol{x}} = f(\boldsymbol{\mu})+\frac{1}{2}\nabla^2\boldsymbol{f}\boldsymbol{P} + \frac{1}{4!}\sum_{i=0}^{2n}w_{i}^{m}\nabla^4\boldsymbol{f} \boldsymbol{\delta}_i^4 + \cdot\cdot\cdot    
\end{equation}
\begin{proof}
\begin{equation}\label{lm2_sp_x_mean_nd_eq_1}
    \centering
    \begin{split}
    \hat{\boldsymbol{x}} = &\sum_{i=0}^{2n} w_{i}^{m} \boldsymbol{x}_i=\\
    &\sum_{i=0}^{2n} w_{i}^{m} (f(\boldsymbol{\mu} + \boldsymbol{\delta}_i) = \\ 
    & \sum_{i=0}^{2n}w_{i}^{m} [f(\boldsymbol{\mu})+\nabla\boldsymbol{f}\boldsymbol{\delta}_i + \frac{\nabla^2\boldsymbol{f} \boldsymbol{\delta}_i^2}{2!}+\frac{\nabla^3\boldsymbol{f}\boldsymbol{\delta}_i^3}{3!}+\cdot\cdot\cdot]=\\
    & \sum_{i=0}^{2n}w_{i}^{m}f(\boldsymbol{\mu})+\sum_{i=0}^{2n}w_{i}^{m}\nabla\boldsymbol{f}\boldsymbol{\delta}_i + \frac{1}{2!}\sum_{i=0}^{2n}w_{i}^{m}\nabla^2\boldsymbol{f} \boldsymbol{\delta}_i^2+\\ &\quad\quad\quad\frac{1}{4!}\sum_{i=0}^{2n}w_{i}^{m}\nabla^4\boldsymbol{f} \boldsymbol{\delta}_i^4 +\cdot\cdot\cdot=\\
    & f(\boldsymbol{\mu})\sum_{i=0}^{2n}w_{i}^{m}+\nabla\boldsymbol{f}\sum_{i=0}^{2n}w_{i}^{m}\boldsymbol{\delta}_i+\frac{1}{2!}\nabla^2\boldsymbol{f}\sum_{i=0}^{2n}w_{i}^{m} \boldsymbol{\delta}_i\boldsymbol{\delta|}_i^T+\\ &\quad\quad\quad\frac{1} {4!}\sum_{i=0}^{2n}w_{i}^{m}\nabla^4\boldsymbol{f} \boldsymbol{\delta}_i^4 +\cdot\cdot\cdot   
    \end{split}
\end{equation}
It follows from Theorems \ref{Theorem1}--\ref{Theorem3} that:
\begin{equation}\label{lm2_sp_x_prop_nd_eq}
\centering
\begin{split}
    & \sum_{i=0}^{2n} w_{i}^{m} = 1 \\
    & \sum_{i=0}^{2n} w_{i}^{m}\mathbf{\boldsymbol{\delta}}_i = 0 \\
    & \sum_{i=0}^{2n} w_{i}^{m}\mathbf{\boldsymbol{\delta}}_i(\mathbf{\boldsymbol{\delta}}_i)^T = \boldsymbol{\mathbf{P}}_{{k}{|}{k}}
\end{split}
\end{equation}
Substituting \eqref{lm2_sp_x_prop_nd_eq} into \eqref{lm2_sp_x_mean_nd_eq_1} gives \eqref{lemma_2_msukf_x_mean_nd_eq}.
\end{proof}
\end{lemma}
\begin{theorem}\label{Theorem4}
The MS-UKF propagated estimated mean $\hat{\boldsymbol{x}}$ is accurate up to the second order and if the deviations are Gaussian, it is accurate up to the third order.
\begin{proof}
The true mean is given by:
\begin{equation}\label{th2_true_mean_eq}
\centering
    E[\boldsymbol{f}(\boldsymbol{\mu+\delta})]= \boldsymbol{f}(\boldsymbol{\mu})+\frac{1}{2}\nabla^2\boldsymbol{f}\boldsymbol{P}+\frac{1}{3!}\nabla^3\boldsymbol{f} E[\boldsymbol{\delta}^3]+\cdot\cdot\cdot
\end{equation}
The difference between the MS-UKF mean \eqref{lemma_2_msukf_x_mean_nd_eq} and the true mean \eqref{th2_true_mean_eq} is:
\begin{equation}\label{th2_msukf_mean_eq_1}
\centering
\begin{split}
    & \hat{\boldsymbol{x}}-E[\boldsymbol{f}(\boldsymbol{\mu+\delta})] =\\
    & f(\boldsymbol{\mu})+\frac{1}{2}\nabla^2\boldsymbol{f}\boldsymbol{P} + \frac{1}{4!}\sum_{i=0}^{2n}w_{i}^{m}\nabla^4\boldsymbol{f}\boldsymbol{\delta}_i^4 + \cdot\cdot\cdot-\\
    & \boldsymbol{f}(\boldsymbol{\mu})+\frac{1}{2}\nabla^2\boldsymbol{f}\boldsymbol{P}+\frac{1}{3!}\nabla^3\boldsymbol{f} E[\boldsymbol{\delta}^3]+\cdot\cdot\cdot =\\
    & \frac{1}{4!}\sum_{i=0}^{2n}w_{i}^{m}\nabla^4\boldsymbol{f}\boldsymbol{\delta}_i^4 + \cdot\cdot\cdot - \frac{1}{3!}\nabla^3\boldsymbol{f}E[\boldsymbol{\delta}^3]+\cdot\cdot\cdot \in O(\|\boldsymbol{\delta}^3\|)
\end{split}
\end{equation}
If the deviations are Gaussian (symmetric), odd moments are zero, so the MS-UKF estimated mean accuracy is up to the third order.
\end{proof}
\end{theorem}
Theorem \ref{Theorem4} shows that the MS-UKF and the standard UKF share the same level of accuracy in term of mean estimation.

\section{}\label{appendix_B_sec}
In this section we show that the estimated covariance accuracy of the MS-UKF is up to second order for a zero mean random variable and up to the third order for Gaussian distribution. The Taylor expansion of $f(\boldsymbol{x})$ around the mean $\boldsymbol{\mu}$ and the true mean $E[f(\boldsymbol{x})]$ are detailed in \eqref{true_taylor_exp_eq} and \eqref{true_mean_eq}. The true covariance is:
\begin{equation}\label{true_cov__eq}
    \centering
    \begin{split}
    \boldsymbol{P}_{true}=& E[(\boldsymbol{f}(\boldsymbol{\mu+\delta}) - E[f(\boldsymbol{x})])(\boldsymbol{f}(\boldsymbol{\mu+\delta}) - E[f(\boldsymbol{x})])^T] =\\
    & \quad \nabla\boldsymbol{f}P(\nabla\boldsymbol{f})^T + O(\|\boldsymbol{\delta}\|^3)
    \end{split}
\end{equation}
\begin{lemma}\label{lemma3}
Let $w^m_0$ and $w^c_0$ be as defined in \eqref{w_m_0_nd_eq} and \eqref{w_c_0_nd_eq}, then:
\begin{equation}\label{sp_cov_w_0_eq}
    \centering
    \begin{split}
    &w_0^c[(f(\boldsymbol{\mu}+\boldsymbol{\delta}_0) - \hat{\boldsymbol{x}})(f(\boldsymbol{\mu}+\boldsymbol{\delta}_0) - \hat{\boldsymbol{x}})^T] =\\ 
    &\quad w_0^m[(f(\boldsymbol{\mu}+\boldsymbol{\delta}_0) - \hat{\boldsymbol{x}})(f(\boldsymbol{\mu}+\boldsymbol{\delta}_0) - \hat{\boldsymbol{x}})^T] + O(\|\boldsymbol{\delta}^4\|)
    \end{split}
\end{equation}
We used here $\boldsymbol{\delta}$ to cover all $\boldsymbol{\delta}_i$ constructed in $\hat{\boldsymbol{x}}$ as described in \eqref{lemma_2_msukf_x_mean_nd_eq}.
\begin{proof}
From \eqref{lemma_2_msukf_x_mean_nd_eq} and since $\delta_0=0$ we obtain:
\begin{equation}\label{lemma3_proof_cov_sp0_element_eq}
    \centering
    \begin{split}
    &(\hat{\boldsymbol{x}} - f(\boldsymbol{\mu}+\boldsymbol{\delta}_0)) = (\hat{\boldsymbol{x}}-f(\mu))=\\
    &\quad \frac{1}{2}\nabla^2\boldsymbol{f}\boldsymbol{P} + \frac{1}{4!}\sum_{i=0}^{2n}w_{i}^{m}\nabla^4\boldsymbol{f}\boldsymbol{\delta}_i^4 + \cdot\cdot\cdot 
    \end{split}
\end{equation}
Thus, $(f(\boldsymbol{\mu}+\boldsymbol{\delta}_0)-\hat{\boldsymbol{x}})\in O(\|\boldsymbol{\delta}^2\|)$.\\
Returning to \eqref{sp_cov_w_0_eq}, we now have:,
\begin{equation}\label{lemma3_proof_eq}
    \centering
    \begin{split}
    &w_0^c[(f(\boldsymbol{\mu}+\boldsymbol{\delta}_0) - \hat{\boldsymbol{x}})(f(\boldsymbol{\mu}+\boldsymbol{\delta}_0) - \hat{\boldsymbol{x}})^T] =\\ 
    &\quad w_0^m[(f(\boldsymbol{\mu}+\boldsymbol{\delta}_0) - \hat{\boldsymbol{x}})(f(\boldsymbol{\mu}+\boldsymbol{\delta}_0) - \hat{\boldsymbol{x}})^T]+\\
    &\quad\gamma[(f(\boldsymbol{\mu}+\boldsymbol{\delta}_0) - \hat{\boldsymbol{x}})(f(\boldsymbol{\mu}+\boldsymbol{\delta}_0) - \hat{\boldsymbol{x}})^T]=\\
    &\quad w_0^m[(f(\boldsymbol{\mu}+\boldsymbol{\delta}_0) - \hat{\boldsymbol{x}})(f(\boldsymbol{\mu}+\boldsymbol{\delta}_0)-\hat{\boldsymbol{x}})^T]+O(\|\boldsymbol{\delta}^4\|)
    \end{split}
\end{equation}
\end{proof}
\end{lemma}

\begin{lemma}\label{lemma4}
Let, $w^c_i, i=0,...,2n$ be as defined in \eqref{w_i_nd_eq},\eqref{w_n_i_nd_eq} and \eqref{w_c_0_nd_eq}, Therefore, the update estimated covariance matrix of the MS-UKF is:
\begin{equation}\label{lemma4_sp_cov_def_eq}
    \centering
    \begin{split}
    &\sum_{i=0}^{2n} w_{i}^{c}[(f(\boldsymbol{\mu}+\boldsymbol{\delta}_i) - \hat{\boldsymbol{x}})(f(\boldsymbol{\mu}+\boldsymbol{\delta}_i) - \hat{\boldsymbol{x}})^T] =\\
    &\quad \nabla f\boldsymbol{P}(\nabla f)^T + O(\|\boldsymbol{\delta}^3\|)
    \end{split}
\end{equation}
where $\boldsymbol{\delta}$ covers all $\boldsymbol{\delta}_i$ constructed in $\hat{\boldsymbol{x}}$ as described in \eqref{lemma_2_msukf_x_mean_nd_eq}
\begin{proof}
Lemma \ref{lemma3} and the definition of $w^m_i, i=0,...2$ given in \eqref{w_m_0_nd_eq} and \eqref{w_i_nd_eq}, imply:
\begin{equation}\label{sp_cov_lemma4_proof_1_eq}
    \centering
    \begin{split}
    & \sum_{i=0}^{2n} w_{i}^{c}[(f(\boldsymbol{\mu}+\boldsymbol{\delta}_i) - \hat{\boldsymbol{x}})(f(\boldsymbol{\mu}+\boldsymbol{\delta}_i) - \hat{\boldsymbol{x}})^T] =\\
    & \sum_{i=0}^{2n} w_{i}^{m}[(f(\boldsymbol{\mu}+\boldsymbol{\delta}_i) - \hat{\boldsymbol{x}})(f(\boldsymbol{\mu}+\boldsymbol{\delta}_i)- \hat{\boldsymbol{x}})^T] + O(\|\boldsymbol{\delta}^4\|)
    \end{split}
\end{equation}
Using \eqref{true_taylor_exp_eq} and \eqref{lemma_2_msukf_x_mean_nd_eq}, we define:
\begin{equation}\label{sp_cov_lemma4_proof_2_eq}
    \centering
    \begin{split}
    & \tilde{\boldsymbol{x}}_i = f(\boldsymbol{\mu}+\boldsymbol{\delta}_i) - \hat{\boldsymbol{x}} = \\
    & \quad f(\boldsymbol{\mu})+\nabla\boldsymbol{f}\boldsymbol{\delta}_i + \frac{\nabla^2\boldsymbol{f}\boldsymbol{\delta}_i^2}{2!}+\frac{\nabla^3\boldsymbol{f}\boldsymbol{\delta}_i^3}{3!}+O(\|\boldsymbol{\delta}_i^4\|) - \hat{\boldsymbol{x}}=\\
    & \quad \nabla\boldsymbol{f}\boldsymbol{\delta}_i + \frac{\nabla^2\boldsymbol{f}\boldsymbol{\delta}_i^2}{2!}+\frac{\nabla^3\boldsymbol{f}\boldsymbol{\delta}_i^3}{3!} - \frac{1}{2}\nabla^2\boldsymbol{f}P +O(\|\boldsymbol{\delta}^4\|)
    \end{split}    
\end{equation}
where $\boldsymbol{\delta}$ covers all $\boldsymbol{\delta}_i$ constructed in $\tilde{\boldsymbol{x}}$. Using Lemma \ref{lemma3} and Theorem \ref{Theorem3}, we obtain:
\begin{equation}\label{sp_cov_lemma4_proof_3_eq}
    \centering
    \begin{split}
    & \sum_{i=0}^{2n} w_{i}^{c}[\tilde{\boldsymbol{x}}_i\tilde{\boldsymbol{x_i}}^T] =\\
    & \sum_{i=0}^{2n} w_{i}^{m}[\tilde{\boldsymbol{x}}_i\tilde{\boldsymbol{x_i}}^T] + O(\|\boldsymbol{\delta}^4\|)=\\
    & \sum_{i=0}^{2n} w_{i}^{m}(\nabla f\boldsymbol{\delta}_i)(\nabla f\boldsymbol{\delta}_i)^T + O(\|\boldsymbol{\delta}^3\|)=\\
    & \sum_{i=0}^{2n} w_{i}^{m}\nabla f\boldsymbol{\delta}_i\boldsymbol{\delta}_i^T(\nabla f)^T + O(\|\boldsymbol{\delta}^3\|)=\\
    & \nabla f(\sum_{i=0}^{2n} w_{i}^{m}\boldsymbol{\delta}_i\boldsymbol{\delta}_i^T)(\nabla f)^T + O(\|\boldsymbol{\delta}^3\|)=\\
    & \nabla f\boldsymbol{P}(\nabla f)^T + O(\|\boldsymbol{\delta}^3\|)
    \end{split}
\end{equation}
\end{proof}
\end{lemma}
\begin{theorem}\label{Theorem5}
MS-UKF propagated covariance is accurate up to the second order and if the deviations are Gaussian, it is accurate up to the third order.
\begin{proof}
From Lemma \ref{lemma4} and \eqref{true_cov__eq} we obtain:
\begin{equation}\label{theorem5_proof_1_eq}
    \centering
    \begin{split}
    & \boldsymbol{P}_{true} - \sum_{i=0}^{2n} w_{i}^{c}[\tilde{\boldsymbol{x}}_i\tilde{\boldsymbol{x}}_i^T] =\\
    & =\nabla f\boldsymbol{P}(\nabla f)^T - \nabla fP(\nabla f)^T + O(\|\boldsymbol{\delta}^3\|)
    \end{split}
\end{equation}
Thus,
\begin{equation}\label{theorem5_proof_2_eq}
    \centering
    \begin{split}
    & P_{true} - \sum_{i=0}^{2n} w_{i}^{c}[\tilde{\boldsymbol{x}}_i(\tilde{\boldsymbol{x_i}})^T] \in O(\|\boldsymbol{\delta}^3\|)
    \end{split}
\end{equation}
and if the distribution is Gaussian, odd moments are zero so the accuracy is up to the third order.
\end{proof}
\end{theorem}
The above shows that the MS-UKF and the standard UKF share the same theoretical level of accuracy regarding the propagated covariance estimation.

\section{}\label{appendix_C_sec}
In this section we give a sufficient condition for the MS-UKF time update covariance matrix to be positive semi-definite.\\
Define:
\begin{equation}\label{AppC_m_eq}
    \centering
    M=\sum_{i=1}^{2n}w^m_i = \sum_{i=1}^{2n}w^c_i=\sum_{i=1}^{2n}\frac{1}{2\Lambda_i}=\sum_{i=1}^{n}\frac{1}{\Lambda_i}
\end{equation}
From \eqref{eq:newx}, denote:
\begin{equation}\label{AppC_Delta_i_eq}
    \centering
    \boldsymbol{\Delta}_i = \boldsymbol{\delta x}_{i,{{k+1}{|}{k}}}, \textbf{ } i=0,...,2n
\end{equation}
We bring here two lemmas that will be useful for the next theorem:
\begin{lemma}\label{lemma5}
\begin{equation}\label{AppC_Delta_separate_3_1_eq}
    \centering
    -(1-M)\boldsymbol{v}^{T}\boldsymbol{\Delta}_0 = \sum_{i=1}^{n}\frac{\boldsymbol{v}^{T}\boldsymbol{\Delta}_i + \boldsymbol{v}^{T}\boldsymbol{\Delta}_{n+1}}{2\Lambda_i}
\end{equation}
where $M, \boldsymbol{\Delta}_i, \Lambda_i$ are defined in \eqref{AppC_m_eq}, \eqref{AppC_Delta_i_eq}, \eqref{Lambda_nd_eq}, respectively, and $\boldsymbol{v} \in \mathbb{R}^n$ is an arbitrary vector.
\begin{proof}
From \eqref{theorem_1_sum_w_1_eq}
\begin{equation}\label{AppC_sum_w_nd_is_1_eq}
    \centering
    \sum_{i=0}^{2n} w_{i}^{m} = 1
\end{equation}
Thus,
\begin{equation}\label{AppC_dev_sum_Delta_0_eq}
    \centering
    \begin{split}
    & \sum_{i=0}^{2n} w_{i}^{m}\boldsymbol{\Delta}_i = \sum_{i=0}^{2n} w_{i}^{m}\boldsymbol{\delta x}_{i,{{k+1}{|}{k}}} =\\
    & \sum_{i=0}^{2n} w_{i}^{m}\left[\boldsymbol{x}_{i,{{k+1}{|}{k}}}-\hat{\boldsymbol{x}}_{{k+1}{|}{k}}\right]=\\
    & \left(\sum_{i=0}^{2n} w_{i}^{m}\boldsymbol{x}_{i,{{k+1}{|}{k}}}\right)-\hat{\boldsymbol{x}}_{{k+1}{|}{k}}=\\
    & \hat{\boldsymbol{x}}_{{k+1}{|}{k}} - \hat{\boldsymbol{x}}_{{k+1}{|}{k}} = 0
    \end{split}
\end{equation}
Leading to
\begin{equation}\label{AppC_sum_Delta_0_eq}
    \centering
    \begin{split}
    \sum_{i=0}^{2n} w_{i}^{m}\boldsymbol{\Delta}_i = 0
    \end{split}
\end{equation}
Separating the first element in the sum and writing the weights explicitly, noticing from \eqref{w_m_0_nd_eq} and \eqref{AppC_m_eq} that $w_0^m = 1-M$, we get:
\begin{equation}\label{AppC_Delta_separate_1_eq}
    \centering
    (1-M)\boldsymbol{\Delta}_0 + \sum_{i=1}^{n}\frac{\boldsymbol{\Delta}_i + \boldsymbol{\Delta}_{n+1}}{2\Lambda_i} = 0
\end{equation}
or
\begin{equation}\label{AppC_Delta_separate_2_eq}
    \centering
    -(1-M)\boldsymbol{\Delta}_0 = \sum_{i=1}^{n}\frac{\boldsymbol{\Delta}_i + \boldsymbol{\Delta}_{n+1}}{2\Lambda_i}
\end{equation}
Multiplying both sides by $\boldsymbol{v}^{T}$ from the left, results in:
\begin{equation}\label{AppC_Delta_separate_3_2_eq}
    \centering
    -(1-M)\boldsymbol{v}^{T}\boldsymbol{\Delta}_0 = \sum_{i=1}^{n}\frac{\boldsymbol{v}^{T}\boldsymbol{\Delta}_i + \boldsymbol{v}^{T}\boldsymbol{\Delta}_{n+1}}{2\Lambda_i}
\end{equation}
\end{proof}
\end{lemma}

\begin{lemma}\label{lemma6}
\begin{equation}\label{lemma6_lower_bound_eq}
    \centering
    \sum_{i=1}^{n}\frac{(\boldsymbol{v}^{T}\boldsymbol{\Delta}_i)^2+(\boldsymbol{v}^{T}\boldsymbol{\Delta}_{n+1})^2}{2\Lambda_i} \geq \frac{(1-M)^2}{M}(\boldsymbol{v}^{T}\boldsymbol{\Delta}_0)^2
\end{equation}
where $\boldsymbol{v} \in \mathbb{R}^n$ is an arbitrary vector.
\begin{proof}
From Cauchy–Schwarz inequality we get:
\begin{equation}\label{AppC_Cauchy_Scharz_ineq_res_1_eq}
    \centering
    \begin{split}
    & \left(\sum_{i=1}^{2n}\left(\sqrt{w^c_i}\right)\left(\sqrt{w^c_i}\boldsymbol{v}^{T}\boldsymbol{\Delta}_i\right)\right)^2 \\
    & \leq \sum_{i=1}^{2n}\left(\sqrt{w^c_i}\right)^2 \sum_{i=1}^{2n}\left(\sqrt{w^c_i}\boldsymbol{v}^{T}\boldsymbol{\Delta}_i\right)^2
    \end{split}
\end{equation}
Note that $\boldsymbol{\Delta}_i\in \mathbb{R}^n$, thus, $\boldsymbol{v}^{T}\boldsymbol{\Delta}_i \in \mathbb{R}$. Then, \eqref{AppC_Cauchy_Scharz_ineq_res_1_eq} is rewritten as:
\begin{equation}\label{AppC_Cauchy_Scharz_ineq_res_2_eq}
    \centering
    \begin{split}
    & \left(\sum_{i=1}^{2n}w^c_i\boldsymbol{v}^{T}\boldsymbol{\Delta}_i\right)^2 \leq \sum_{i=1}^{2n}w^c_i \sum_{i=1}^{2n}w^c_i(\boldsymbol{v}^{T}\boldsymbol{\Delta}_i)^2
    \end{split}
\end{equation}
Since $w^c_i = w^c_{n+i}=\frac{1}{2\Lambda_i}$ for $i=1,...n$, from \eqref{AppC_m_eq} we can write:
\begin{equation}\label{AppC_Cauchy_Scharz_ineq_res_3_1_eq}
    \centering
    \begin{split}
    & \sum_{i=1}^{2n}w^c_i \sum_{i=1}^{2n}w^c_i(\boldsymbol{v}^{T}\boldsymbol{\Delta}_i)^2 = \\
    & M \sum_{i=1}^{n}w^c_i[(\boldsymbol{v}^{T}\boldsymbol{\Delta}_i)^2 + (\boldsymbol{v}^{T}\boldsymbol{\Delta}_{n+i})^2]=\\
    & M \sum_{i=1}^{n}\frac{(\boldsymbol{v}^{T}\boldsymbol{\Delta}_i)^2 + (\boldsymbol{v}^{T}\boldsymbol{\Delta}_{n+i})^2}{2\Lambda_i}\\
    \end{split}
\end{equation}
Looking on the left side of \eqref{AppC_Cauchy_Scharz_ineq_res_2_eq}, using Lemma \ref{lemma5} and writing the weights explicitly yields:
\begin{equation}\label{AppC_Cauchy_Scharz_left_side_eq}
    \centering
    \begin{split}
    & \left(\sum_{i=1}^{2n}w^c_i\boldsymbol{v}^{T}\boldsymbol{\Delta}_i\right)^2 =\\
    & \left(\sum_{i=1}^{n}\frac{\boldsymbol{v}^{T}\boldsymbol{\Delta}_i + \boldsymbol{v}^{T}\boldsymbol{\Delta}_{n+1}}{2\Lambda_i}\right)^2\\
    & = (-(1-M)\boldsymbol{v}^{T}\boldsymbol{\Delta}_0)^2=(1-M)^2(\boldsymbol{v}^{T}\boldsymbol{\Delta}_0)^2        
    \end{split}
\end{equation}
Thus:
\begin{equation}\label{AppC_Cauchy_Scharz_ineq_res_3_2_eq}
    \centering
    (1-M)^2(\boldsymbol{v}^{T}\boldsymbol{\Delta}_0)^2 \leq M \sum_{i=1}^{n}\frac{(\boldsymbol{v}^{T}\boldsymbol{\Delta}_i)^2 + (\boldsymbol{v}^{T}\boldsymbol{\Delta}_{n+i})^2}{2\Lambda_i}
\end{equation}
From \eqref{Lambda_nd_eq} and \eqref{w_i_positive_cond_nd_eq} $\Lambda_i > 0$, so $M > 0$ leading to the following lower bound:
\begin{equation}\label{AppC_Cauchy_Scharz_ineq_res_4_eq}
    \centering
    \frac{(1-M)^2}{M}(\boldsymbol{v}^{T}\boldsymbol{\Delta}_0)^2 \leq \sum_{i=1}^{n}\frac{(\boldsymbol{v}^{T}\boldsymbol{\Delta}_i)^2 + (\boldsymbol{v}^{T}\boldsymbol{\Delta}_{n+i})^2}{2\Lambda_i}
\end{equation}
\end{proof}
\end{lemma}

\begin{theorem}\label{Theorem6}
A sufficient condition for the MS-UKF propagated covariance matrix $\boldsymbol{\mathbf{P}}_{{k+1}{|}{k}} \in \mathbb{R}^{n \times n}$ at time step $k+1$ ($k \geq 0$) to be positive semi-definite is:
\begin{equation}\label{AppC_th6_psd_criteria_eq}
    \centering
    \left( \sum_{i=1}^{n}\frac{1}{\alpha^2_i (n+\kappa_i)} \right)^{-1}-(\sqrt[n]{\alpha_1} \cdots \sqrt[n]{\alpha_n})^2 + \beta \geq 0
\end{equation}
where $n > 0$ is the size of the state vector.
\begin{proof}
From the definition of a real positive semi-definite matrix, given a vector $\boldsymbol{v} \in \mathbb{R}^n$, we want to find a sufficient condition such that:
\begin{equation}\label{AppC_psd_eq}
    \centering
    \boldsymbol{v}^{T}\boldsymbol{\mathbf{P}}_{{k+1}{|}{k}}\boldsymbol{v} \geq 0
\end{equation}
where $\boldsymbol{\mathbf{P}}_{{k+1}{|}{k}}$ is given in \eqref{tu_p_eq}. Therefore,
\begin{equation}\label{AppC_psd_detailed_eq}
    \centering
    \begin{split}
    & \boldsymbol{v}^{T}\boldsymbol{\mathbf{P}}_{{k+1}{|}{k}}\boldsymbol{v} = 
    \boldsymbol{v}^{T}(\sum_{i=0}^{2n}w_{i}^{c}{\boldsymbol{\delta x}_{i,{{k+1}{|}{k}}} \cdot {\boldsymbol{\delta x}_{i,{{k+1}{|}{k}}}}^T} + \boldsymbol{\mathbf{Q}}_{k})\boldsymbol{v}=\\
    & \boldsymbol{v}^{T}(\sum_{i=0}^{2n}w_{i}^{c}{\boldsymbol{\Delta}_i \cdot {\boldsymbol{\Delta}_i}^T} + \boldsymbol{\mathbf{Q}}_{k})\boldsymbol{v}=\\
    & \boldsymbol{v}^{T}(1-M+\gamma)\boldsymbol{\Delta}_0\boldsymbol{\Delta}_0^T\boldsymbol{v}+\boldsymbol{v}^{T}(\sum_{i=1}^{2n}\frac{\boldsymbol{\Delta}_i \cdot {\boldsymbol{\Delta}_i}^T}{2\Lambda_i})\boldsymbol{v}+\\ & \boldsymbol{v}^{T}\boldsymbol{\mathbf{Q}}_{k}\boldsymbol{v}=\\
    & (1-M+\gamma)(\boldsymbol{v}^{T}\boldsymbol{\Delta}_0)^2+\sum_{i=1}^{2n}\frac{(\boldsymbol{v}^{T}\boldsymbol{\Delta}_i)^2}{2\Lambda_i}+\boldsymbol{v}^{T}\boldsymbol{\mathbf{Q}}_{k}\boldsymbol{v}=\\
    & (1-M+\gamma)(\boldsymbol{v}^{T}\boldsymbol{\Delta}_0)^2+\sum_{i=1}^{n}\frac{(\boldsymbol{v}^{T}\boldsymbol{\Delta}_i)^2+(\boldsymbol{v}^{T}\boldsymbol{\Delta}_{n+1})^2}{2\Lambda_i}+\\
    &\boldsymbol{v}^{T}\boldsymbol{\mathbf{Q}}_{k}\boldsymbol{v}
    \end{split}
\end{equation}
Using Lemma \ref{lemma6}, gives:
\begin{equation}\label{AppC_psd_detailed_1_eq}
    \centering
    \begin{split}
    & \boldsymbol{v}^{T}\boldsymbol{\mathbf{P}}_{{k+1}{|}{k}}\boldsymbol{v} = \\
    & (1-M+\gamma)(\boldsymbol{v}^{T}\boldsymbol{\Delta}_0)^2+\sum_{i=1}^{n}\frac{(\boldsymbol{v}^{T}\boldsymbol{\Delta}_i)^2+(\boldsymbol{v}^{T}\boldsymbol{\Delta}_{n+1})^2}{2\Lambda_i}+ \\ 
    & \boldsymbol{v}^{T}\boldsymbol{\mathbf{Q}}_{k}\boldsymbol{v} \geq (1-M+\gamma)(\boldsymbol{v}^{T}\boldsymbol{\Delta}_0)^2 + \\
    & \frac{(1-M)^2}{M}(\boldsymbol{v}^{T}\boldsymbol{\Delta}_0)^2+\boldsymbol{v}^{T}\boldsymbol{\mathbf{Q}}_{k}\boldsymbol{v} = \\
    & \left(1-M+\gamma + \frac{(1-M)^2}{M}\right)(\boldsymbol{v}^{T}\boldsymbol{\Delta}_0)^2+\boldsymbol{v}^{T}\boldsymbol{\mathbf{Q}}_{k}\boldsymbol{v}
    \end{split}
\end{equation}
As $\mathbf{Q}_{k}$ is a process noise covariance matrix, it is positive semi-definite. Thus, the following condition for $\boldsymbol{\mathbf{P}}_{{k+1}{|}{k}}$ to be positive semi-definite is provided:
\begin{equation}\label{AppC_psd_detailed_2_eq}
    \centering
    0 \leq 1-M+\gamma + \frac{(1-M)^2}{M}=\frac{1}{M}-1+\gamma
\end{equation}
From \eqref{AppC_m_eq}:
\begin{equation}\label{AppC_M_detailed_eq}
    \centering
    M=\sum_{i=1}^{n}\frac{1}{\Lambda_i}=\sum_{i=1}^{n}\frac{1}{\alpha^2_i(n+\kappa_i)}
\end{equation}
and from \eqref{gamma_nd_eq}:
\begin{equation}\label{AppC_gamma_detailed_eq}
    \centering
    \gamma=1 - (\sqrt[n]{\alpha_1} \cdots \sqrt[n]{\alpha_n})^2 + \beta
\end{equation}
Rewriting \eqref{AppC_psd_detailed_2_eq} explicitly, using \eqref{AppC_M_detailed_eq}-\eqref{AppC_gamma_detailed_eq} yields the required condition:
\begin{equation}\label{AppC_psd_criteria_eq}
    \centering
    \left( \sum_{i=1}^{n}\frac{1}{\alpha^2_i (n+\kappa_i)} \right)^{-1}-(\sqrt[n]{\alpha_1} \cdots \sqrt[n]{\alpha_n})^2 + \beta \geq 0
\end{equation}
\end{proof}
\end{theorem}

\end{appendices}

%\section{Disclosure statement}\label{disclosure_statement_sec}
%No potential conflict of interest was reported by the author(s).

%\section{Data availability statement}\label{data_avail_statement_sec}
%Data sharing is not applicable to this article as no new data were created or analyzed in this study.

% ----------------------- End article ----------------------------------

\bibliographystyle{IEEEtran}
\bibliography{pIBio}

\end{document}